\documentclass[useAMS,usenatbib]{mn2e}
\usepackage{amsmath}
\usepackage{epsfig}
\usepackage{txfonts}

\usepackage[pdftex,pdfpagemode={UseOutlines},bookmarks,bookmarksopen,colorlinks,linkcolor={blue},citecolor={blue},urlcolor={red}]{hyperref}
\usepackage{float}
\usepackage{graphicx}
\usepackage{epstopdf}
\usepackage{caption}
\usepackage{subcaption}
\usepackage{natbib}
\usepackage{rotating,longtable,lscape}
\usepackage{multirow}
\usepackage{pifont}
\newcommand{\cmark}{\ding{51}}%

\DeclareCaptionFormat{cont}{#1 #2#3\par}

\title[Time-dependent circumbinary transitability]{Circumbinary planets II - when transits come and go}

\author[Martin]
{\parbox{\textwidth}{David. V. Martin\thanks{E-mail: david.martin@unige.ch}}
\vspace{0.4cm}\\
\parbox{\textwidth}{Observatoire de Gen\`eve, Universit\'e de Gen\`eve, 51 chemin des Maillettes, Sauverny 1290, Switzerland}}

\begin{document}

\date{Accepted . Received}

\pagerange{\pageref{firstpage}--\pageref{lastpage}} \pubyear{2016}

\maketitle

\label{firstpage}

\begin{abstract}

Circumbinary planets are generally more likely to transit than equivalent single-star planets, but practically the geometry and orbital dynamics of circumbinary planets make the chance of observing a transit inherently time-dependent. In this follow-up paper to \citet{martin15}, the time-dependence is probed deeper by analytically calculating when and for how long the binary and planet orbits overlap, allowing for transits to occur. The derived equations are applied to the known transiting circumbinary planets found by {\it Kepler} to predict when future transits will occur, and whether they will be observable by upcoming space telescopes {\it TESS}, {\it CHEOPS} and {\it PLATO}. The majority of these planets spend less than 50\% of their time in a transiting configuration, some less than 20\%. From this it is calculated that the known  {\it Kepler} eclipsing binaries likely  host an additional $\sim 17-30$  circumbinary planets that are similar to the ten published discoveries, and they will ultimately transit some day, potentially during the {\it TESS} and {\it PLATO} surveys.

\end{abstract}


\section{Introduction}
\label{sec:intro}

As early as the 1930's, people marvelled at the then science fiction concept of a planet orbiting two stars - a circumbinary planet \citep{rudaux37}. Astronomers have contemplated their existence and characteristics even before the birth of exoplanet discoveries. \citet{borucki84} noted that photometric searches around eclipsing binaries would have enhanced transit probabilities, which was later expanded upon by \citet{schneider90} and \citet{schneider94}. After the unambiguous discovery of the first transiting circumbinary planet Kepler-16 \citep{doyle11} the field has flourished, leading to an additional ten transiting discoveries (the latest being Kepler-1647 by \citealt{kostov16}) and a wealth of related studies.

For a circumbinary planet to transit it must pass in front of a moving target. This is fundamentally different to a stationary single star, and leads to enhanced transit timing variations \citep{agol05,holman05,armstrong13a} and transit duration variations \citep{kostov14,liu14}. If the planet and binary orbits are coplanar then transits are only possible on eclipsing binaries, but in this case transits are guaranteed once per planet orbit. If the planet and binary orbits are misaligned then transits are still possible, even on non-eclipsing binaries, but there will be gaps in the transit sequence and asymmetries in the transit profiles \citep{martin14}. The picture is further complicated by the rapid orbital dynamics of circumbinary systems, caused by perturbations from the binary. The planet's orbit precesses on an observationally-relevant timescale of years \citep{schneider94,farago10,doolin11,leung13}. Consequently, the state of the planet and binary orbits overlapping on the sky - essential for transits - changes with time, as was seen in the discovery of Kepler-413 \citep{kostov14}. 

Knowing the probability and time-dependence of circumbinary transits has wide implications, including:

\begin{itemize}

\item De-biasing observations to infer properties about circumbinary architectures and abundance \citep{armstrong14,martin14,li16}.
\item Scheduling future photometric observations, both of known-transiting systems and ones found by alternative methods, e.g. eclipse timing variations \citep{schwarz11,schwarz16}, radial velocities \citep{konacki09} and astrometry \citep{sahlmann14}.
\item Predicting discovery yields for transit surveys, like {\it Kepler} in years gone by \citep{martin14} and {\it TESS} and {\it PLATO} to come \citep{li16}.
\item Understanding the fundamental differences between single and binary star systems, and in some cases exploiting the advantageous properties of circumbinary planets \citep{borucki84,schneider94,martin14,martin15,li16}.

\end{itemize}

Being able to calculate transit probabilities analytically is much more efficient than using N-body integrations, and also better illuminates the geometry and orbital dynamics. However, unlike for single stars, it is not trivial to make such calculations for circumbinary planets. To help the analysis, in \citet{martin14} we formally defined the concept of {\it transitability}: a state in which the planet and binary orbits overlap on the sky, meaning that transits are possible but not guaranteed on every passing of the binary orbit. Transit probabilities were calculated numerically using N-body simulations for a handful of example systems, mainly applicable to the {\it Kepler} mission, and shown to be generally higher than for equivalent single-star planets. The next goal has been to calculate {\it analytically} the circumbinary transit probability, for any configuration and over any observing timespan. This task has been split into a series of three papers, of which this present paper constitutes the juicy meat in the circumbinary sandwich:

\begin{enumerate}

\item {\bf \citet{martin15}:} analytic derivation of the probability that a given circumbinary planet would enter transitability at some {\it unspecified} point in time. It was shown numerically that transitability ultimately guarantees transits if you look for long enough, and hence we had derived a {\it time-infinite} transit probability. The numerical work also showed that transits frequently occurred within reasonable timeframes ($\sim$ years). Applied to eclipsing binaries, the analytic work predicts that almost {\it all} circumbinary planets orbiting eclipsing binaries will eventually transit.

\item {\bf This paper:} analytic derivation of the time-dependence of transitability, calculating when planets enter and exit it. By knowing the temporal windows of transitability, it advances the work in \citet{martin15} and improves applicability to astronomers with less than infinite time available.

\item {\bf Future work:} calculation of the efficiency of transitability at producing transits, ultimately, yielding a complete time-dependent transit probability. This task is made difficult by the high sensitivity of the transit sequence to orbital parameters, as explored in \citet{martin14}, and the broad parameter space.
\end{enumerate}

The calculations made in this paper are kept as general as possible, accounting for any binary and planet inclination. Even though planets have only been discovered to date around eclipsing binaries, most binaries do not eclipse. Even though only coplanar planets have been found to date, circumbinary planets have been suggested to become misaligned due to planet-planet scattering \citep{chatterjee08,smullen16} or under the influence of an outer third star \citep{munoz15,martinetal15,hamers16}. 

 Circumbinary transit probabilities have also been analysed in two recent papers. \citet{li16} followed a similar vein to derive a time-dependent transit probability, but only for planets around eclipsing binaries. Their work was also viewed as an extension of \citet{martin15}. The primary purpose was to de-bias the observed trends in circumbinary planets to uncover their architectures. It was done using a Bayesian framework and with more rigour than the earlier study in \citet{martin14}.  \citet{brakensiek16} developed the semi-analytic algorithm CORBITS\footnote{Freely available at \url{https://github.com/jbrakensiek/CORBITS}.} to calculate the transit probability of any pairs of bodies, with primary applications to the {\it Kepler} multi-planet systems and the Solar System. Their algorithm has an orders of magnitude speed increase compared with N-body Monte Carlo simulations. It may be applied to circumbinary planets, but with the caveat the it does not account for precession of the planet's orbital plane, which becomes important for long observing timespans such as the four-year {\it Kepler} mission.

This present paper is sliced up as follows. First, in Sect.~\ref{sec:problem_setup} we setup the circumbinary geometry and orbital dynamics to be used. Following this in Sect.~\ref{sec:derivation} is the analytic calculation of the time-dependence of transitability.  In Sect.~\ref{sec:analysis} we analyse the derived equations and some of the assumptions used. Finally, in Sect.~\ref{sec:applications} we apply our work to the known transiting circumbinary planets discovered by the {\it Kepler} mission and predict transits to be observed by future space missions, before concluding.

%
%
%
%
%
%
%
%
%
%

\section{Problem setup}\label{sec:problem_setup}

\subsection{Geometry}

\begin{figure}  
\begin{center}  
\includegraphics[width=0.5\textwidth]{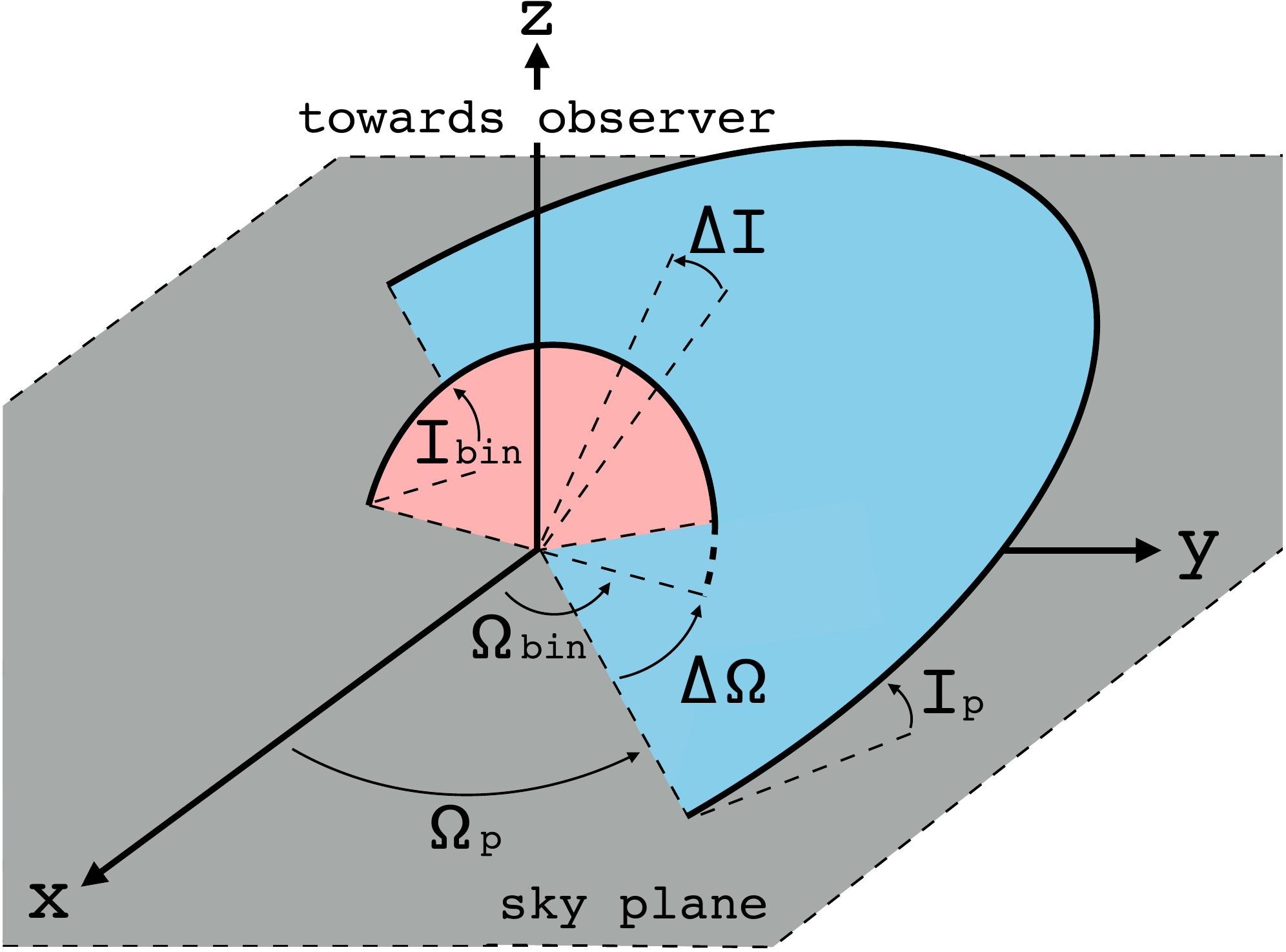}  
\caption{A misaligned circumbinary planet (blue, outer) orbiting a binary star system (pink, inner). The observer is looking down the positive z-axis (from above) so the x-y plane is the plane of the sky. The misalignment between the planet and binary orbits is characterised by the mutual inclination, $\Delta I$ (Eq.~\ref{eq:Delta_I}) and the mutual longitude of the ascending node, $\Delta \Omega$ (Eq.~\ref{eq:Delta_Omega}). Figure has been reproduced from \citet{martin15}.}
\label{fig:Geometry_3D}
\end{center}  
\end{figure} 

\begin{figure}  
\begin{center}  
\includegraphics[width=0.5\textwidth]{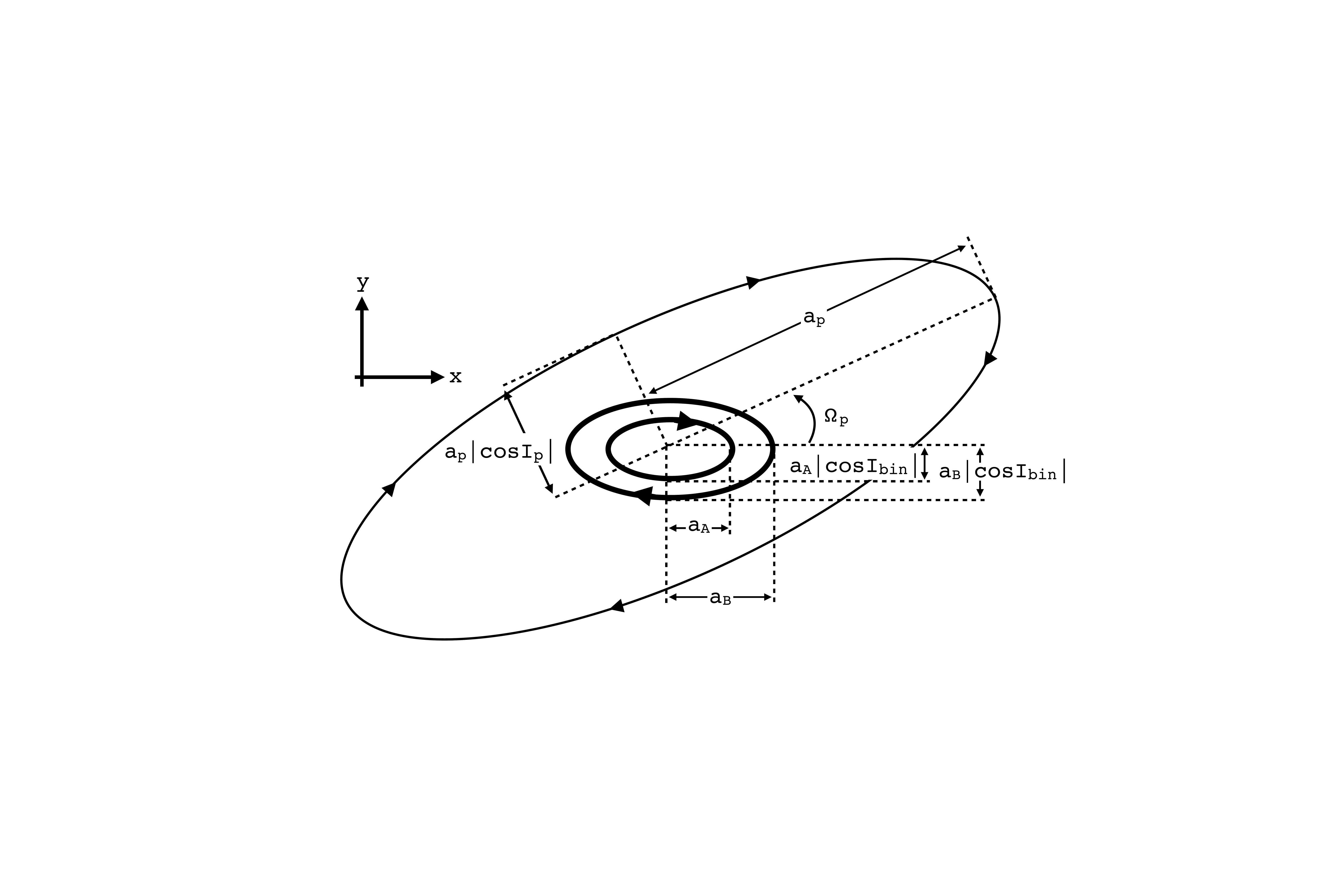}  
\caption{Orbits of a circumbinary planet and binary stars A and B, projected onto the x-y plane of the sky. To simplify the geometry, we have arbitrarily taken $\Omega_{\rm bin}=0$.}
\label{fig:ProjectedEllipses}
\end{center}  
\end{figure} 

\begin{figure*}  
\begin{center}  
	\begin{subfigure}[b]{0.33\textwidth}
		\caption{$I_{\rm bin} = 90^{\circ}$}
		\includegraphics[width=\textwidth]{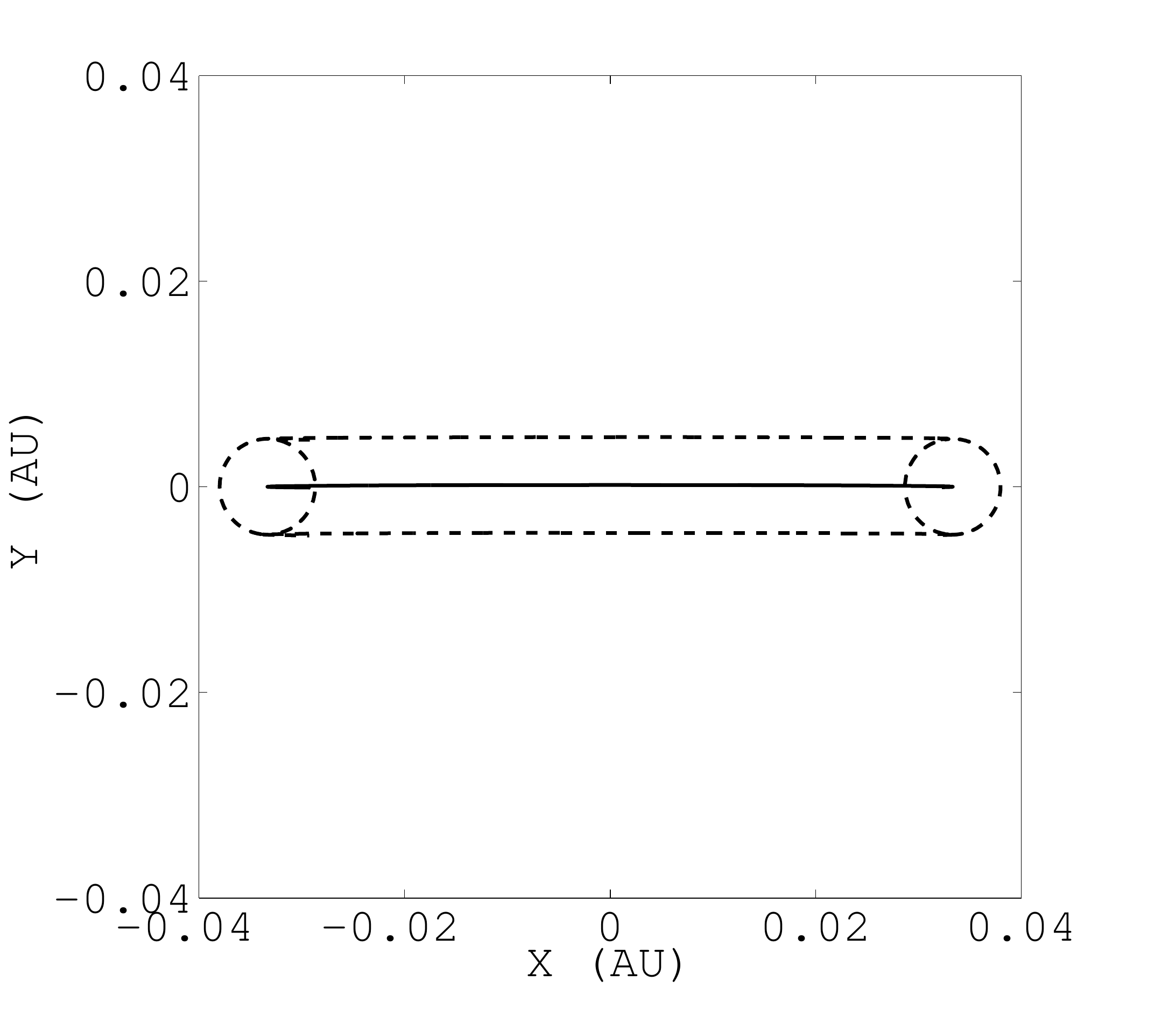}  
	\end{subfigure}
	\begin{subfigure}[b]{0.33\textwidth}
		\caption{$I_{\rm bin} = 88^{\circ}$}
		\includegraphics[width=\textwidth]{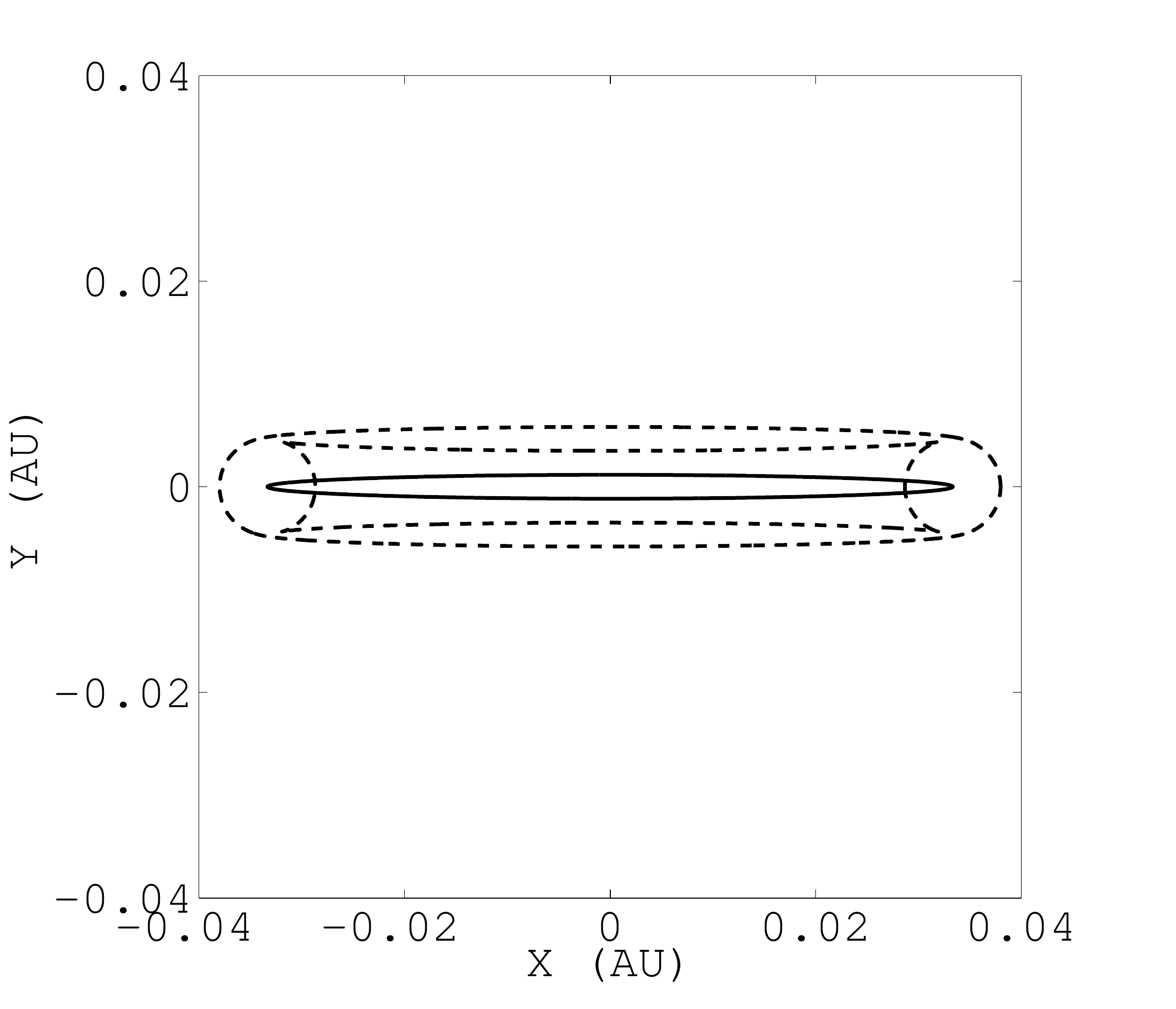}  
	\end{subfigure}
	\begin{subfigure}[b]{0.33\textwidth}
		\caption{$I_{\rm bin} = 80^{\circ}$}
		\includegraphics[width=\textwidth]{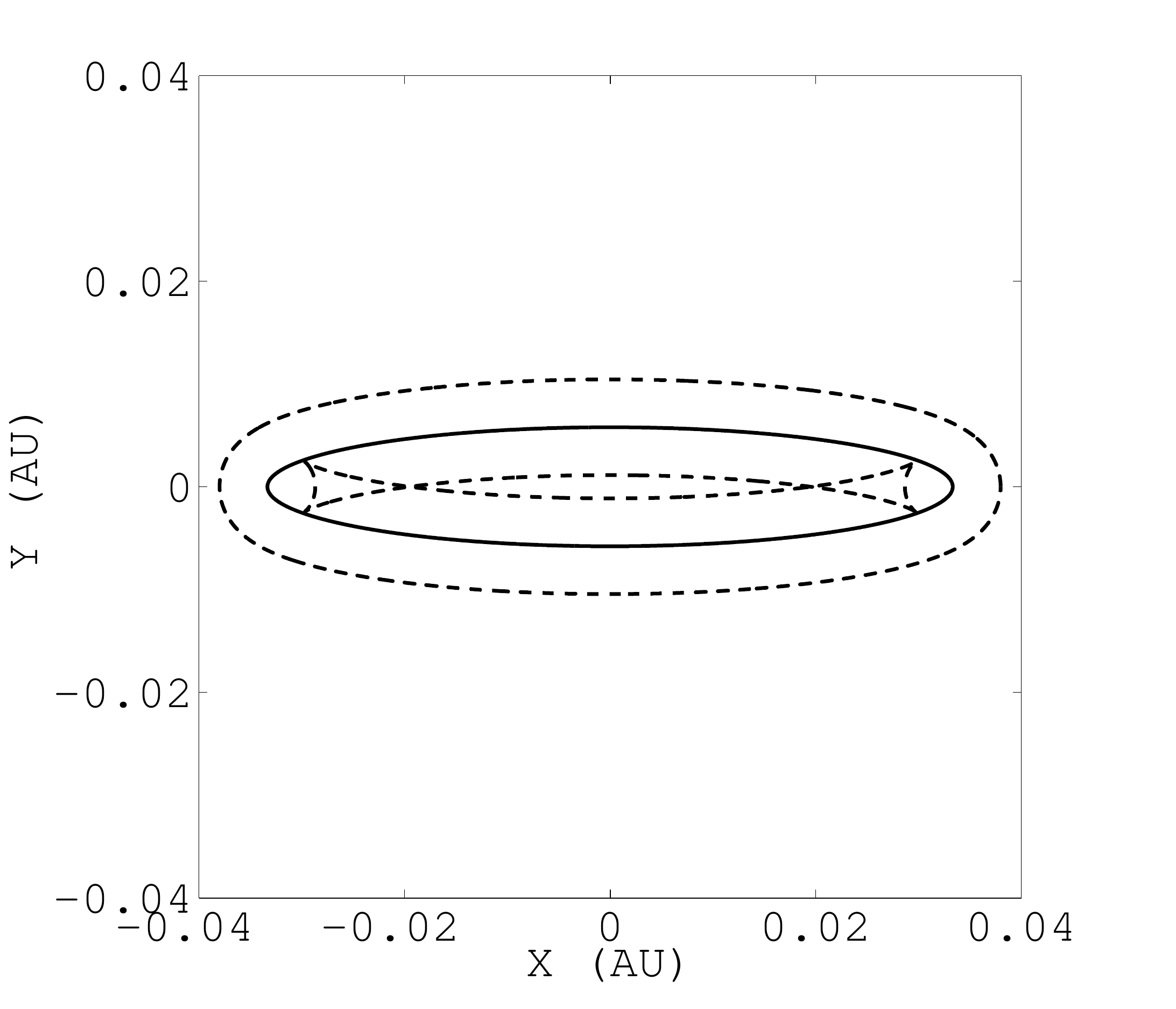}  	
	\end{subfigure}
	\begin{subfigure}[b]{0.33\textwidth}
		\caption{$I_{\rm bin} = 75^{\circ}$}
		\includegraphics[width=\textwidth]{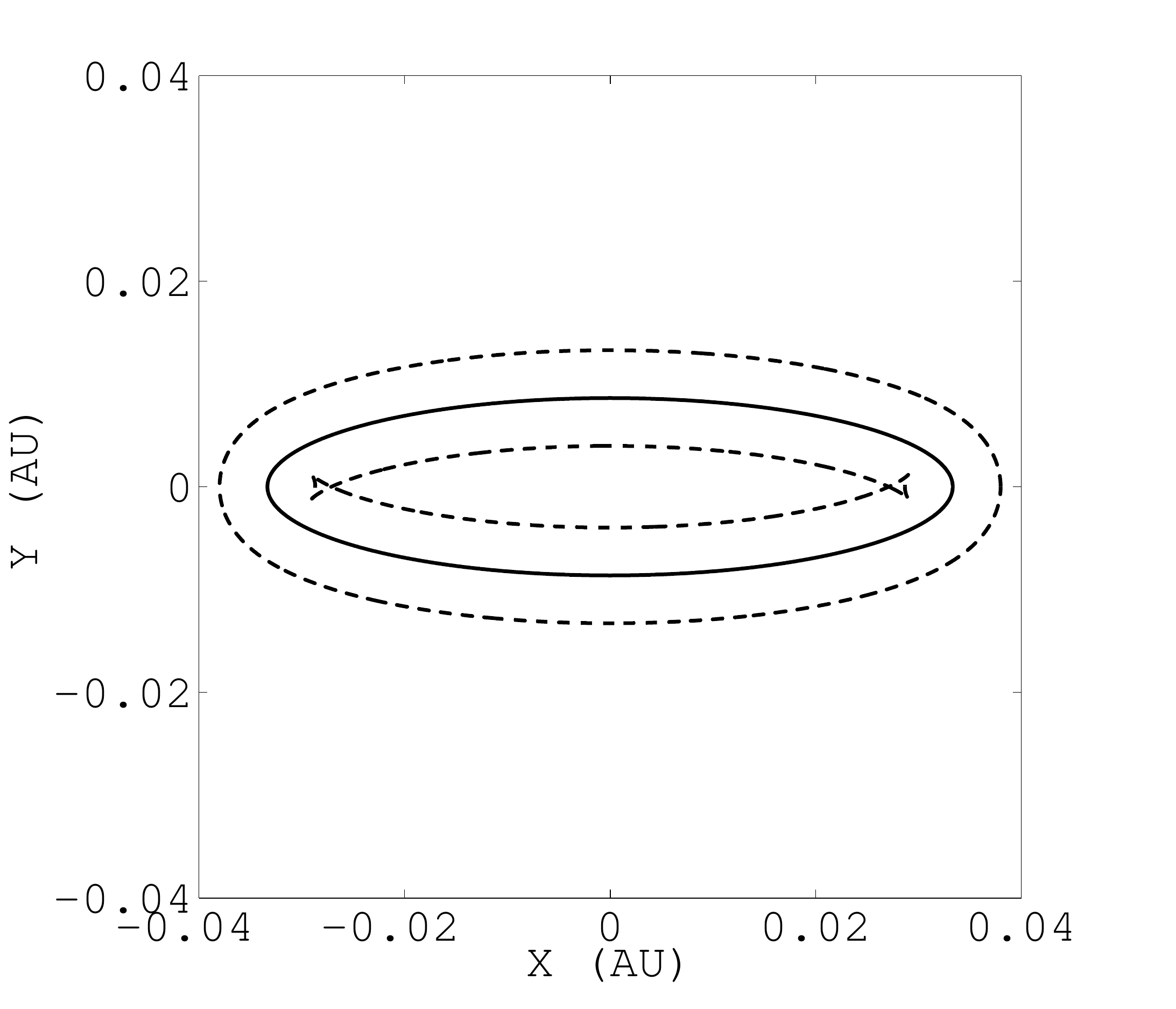}  
	\end{subfigure}
	\begin{subfigure}[b]{0.33\textwidth}
		\caption{$I_{\rm bin} = 70^{\circ}$}
		\includegraphics[width=\textwidth]{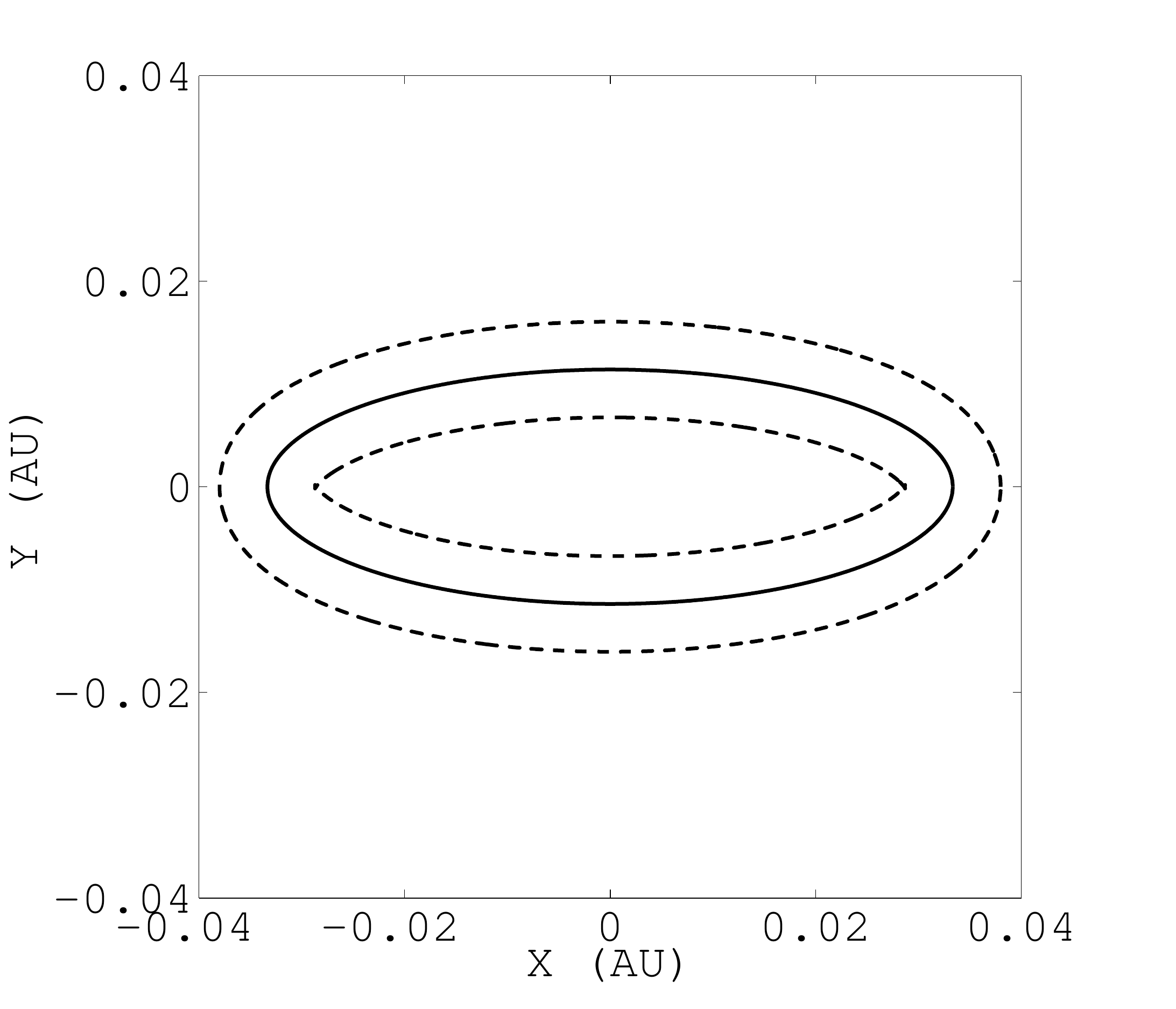}  
	\end{subfigure}
	\begin{subfigure}[b]{0.33\textwidth}
		\caption{$I_{\rm bin} = 0^{\circ}$}
		\includegraphics[width=\textwidth]{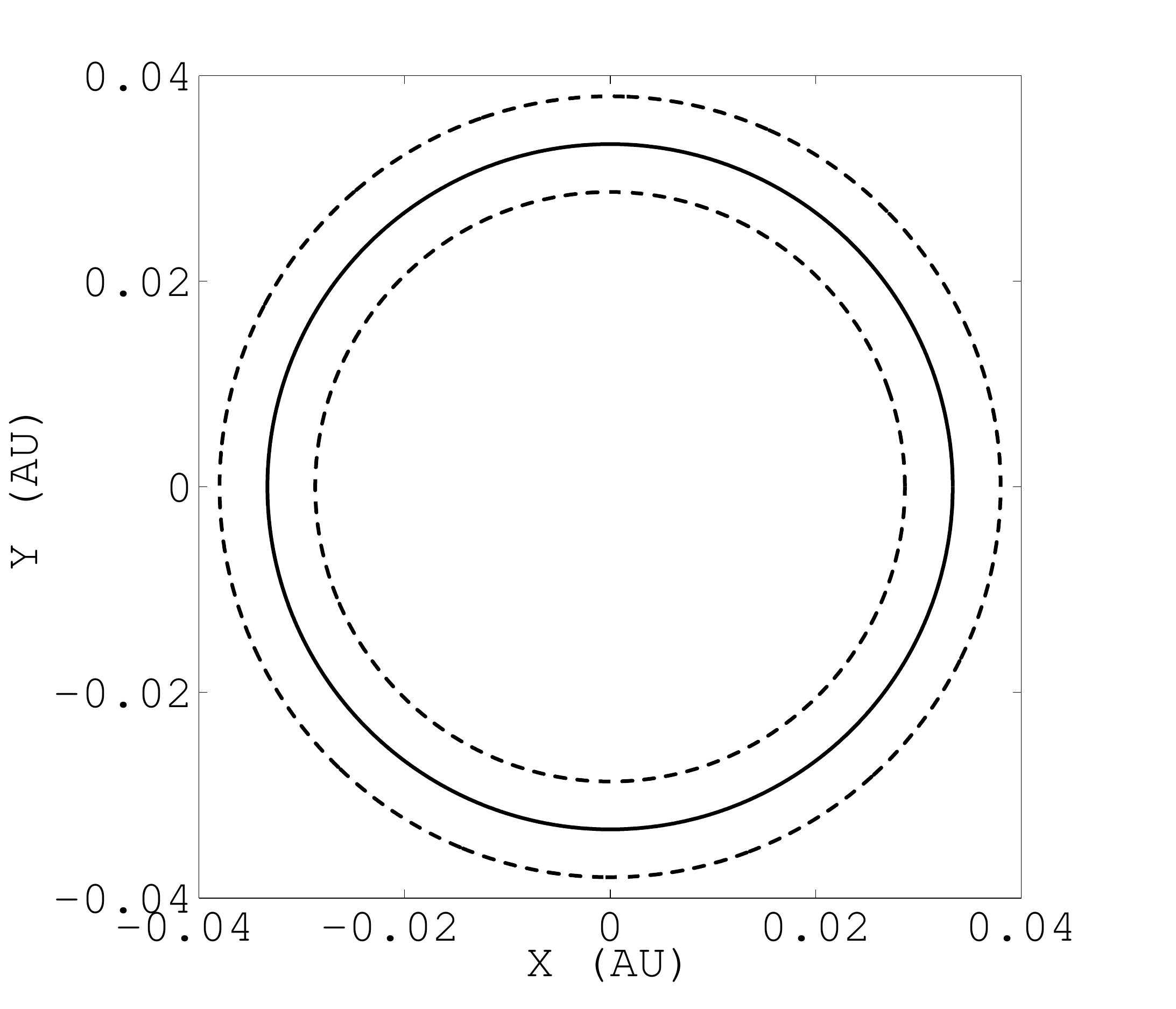}  
	\end{subfigure}
	\caption{Projected primary stellar orbits on the x-y sky plane for $a_{\rm bin} = 0.1$ AU, $R_{\rm A}= 1R_{\odot}$, $M_{\rm A} = 1M_{\odot}$, $M_{\rm B} = 0.5M_{\odot}$ and six different values of $I_{\rm bin}$. The solid line in the middle is the binary ellipse (Eq.~\ref{eq:CoM_curve}). The outer and inner dashed lines show the full extent of the moving stellar disc, using Eqs.~\ref{eq:outer_parallel_curve} and ~\ref{eq:inner_parallel_curve}, respectively. The limiting case of a perfectly edge-on orbit in (a, $I_{\rm bin}= 90^{\circ}$) is described by Eqs.~\ref{eq:outer_curve_90deg} and ~\ref{eq:inner_curve_90deg}, except for the rounded edges that these limiting equations do not account for. The limiting case of a perfectly face-on orbit in (e, $I_{\rm bin} = 0^{\circ}$) is described by Eqs.~\ref{eq:outer_curve_0deg} and ~\ref{eq:inner_curve_0deg}. In (e) $I_{\rm bin}=70^{\circ}$ and this is roughly where the orbital extent starts to look like an annulus.
}\label{fig:stellar_orbit_extent}  
\end{center}  
\end{figure*} 

A circumbinary system is defined by two inner stars, of mass and radius $M_{\rm A,B}$ and $R_{\rm A,B}$, where we use ``A" and ``B" to denote the primary and secondary stars, and an outer planet with mass and radius $M_{\rm p}$ and $R_{\rm p}$. The system is characterised using two Keplerian orbits, one for the binary and one for the planet orbiting the binary's centre of mass, defined by osculating Jacobi elements. The planet and binary orbits are each defined by six orbital elements, which are chosen to be the semi-major axis, $a$, eccentricity, $e$, inclination, $I$, longitude of the ascending node, $\Omega$, argument of periapse, $\omega$ and true longitude, $\theta$. Sometimes instead of $a$ the orbital period is used, which we denote with $T$. A ``p" subscript is used to denote planet quantities. For the binary, we either use ``bin" for quantities general to both stars or ``A" and ``B" for quantities to specific to each individual star.

In Fig.~\ref{fig:Geometry_3D} we illustrate an example circumbinary system where the planetary orbit (blue, outer) is misaligned to the binary orbit (pink, inner) by

\begin{equation}
\label{eq:Delta_I}
\cos \Delta I = \cos \Delta \Omega\sin I_{\rm bin}\sin I_{\rm p} + \cos I_{\rm bin} \cos I_{\rm p},
\end{equation}
where the mutual longitude of the ascending node is

\begin{equation}
\label{eq:Delta_Omega}
\Delta \Omega = \Omega_{\rm bin} - \Omega_{\rm p}.
\end{equation}
In this geometry, an eclipsing binary has $I_{\rm bin}\approx 90^{\circ}$. As an observer, we are only sensitive to $\Delta \Omega$ and not the two individual longitudes, and hence we may arbitrarily take $\Omega_{\rm bin}=0$ for  simplicity.

The planet is assumed to have zero mass, since this outer body has a negligible effect on the orbital dynamics as long as it is roughly within the planetary regime (see \citet{migaszewski11} and \citet{martin16} for more detail). The planet radius is also taken to be zero, since it generally has a negligible effect on the transit probabilities. Finally, for the geometry both binary and planetary orbits are assumed to be circular,  but the effects of including eccentricity are analysed in Sect.~\ref{subsec:eccentricity}.

The primary and secondary star orbits project an ellipse on the $(x,y)$ plane of the sky:

\begin{equation}
\label{eq:binary_orbit}
\frac{x_{\rm A,B}^2}{a_{\rm A,B}^2} + \frac{y_{\rm A,B}^2}{a_{\rm A,B}^2\cos^2I_{\rm bin}} = 1,
\end{equation}
where 

\begin{equation}
\label{eq:stellar_semis}
a_{\rm A} = a_{\rm bin}\frac{M_{\rm B}}{M_{\rm A} + M_{\rm B}} \quad {\rm and} \quad a_{\rm B} = a_{\rm bin}\frac{M_{\rm A}}{M_{\rm A} + M_{\rm B}}
\end{equation}
are the semi-major axes of the two individual stars. The planet similarly projects an ellipse on the sky,

\begin{equation}
\label{eq:planet_orbit}
\frac{\left(x_{\rm p}\cos\Omega_{\rm p} + y_{\rm p}\sin\Omega_{\rm p}\right)^2}{a_{\rm p}^2} + \frac{\left(x_{\rm p}\sin\Omega_{\rm p} - y_{\rm p}\cos\Omega_{\rm p}\right)^2}{a_{\rm p}^2\cos^2 I_{\rm p}} = 1,
\end{equation}
but unlike the binary orbit, the ellipse defining the planetary orbit is rotated anti-clockwise on the plane of the sky by the angle $\Omega_{\rm p}$. 

Equation~\ref{eq:binary_orbit} tracks the motion of the centre of each star, but for transits the motion of the each stellar disc is important. To define that we use offset curves, also known as parallel curves \citep{yates52}. For a parametric curve 

\begin{equation}
\left[\begin{array}{c}x \\y\end{array}\right] = \left[\begin{array}{c}f(\theta) \\ g(\theta)  \end{array}\right],
\end{equation}
where $f(\theta)$ and $g(\theta)$ are some arbitrary functions, the two branches of the offset curve at a distance $k$ are

\begin{equation}
\left[\begin{array}{c}x_{\rm offset} \\y_{\rm offset}\end{array}\right] = \left[\begin{array}{c} f \pm \frac{kg'}{\sqrt{f'^2 + g'^2}}\\  g \mp \frac{kf'}{\sqrt{f'^2 + g'^2}}\end{array}\right]
\end{equation}

To calculate the binary offset curves, first convert Eq.~\ref{eq:binary_orbit} into a parametric equation of the true longitude $\theta_{\rm A,B}$:

\begin{equation}
\label{eq:CoM_curve}
\left[\begin{array}{c}x_{\rm A,B} \\y_{\rm A,B}\end{array}\right] = \left[\begin{array}{c}a_{\rm A,B} \cos \theta_{\rm A,B} \\ a_{\rm A,B} |\cos I_{\rm bin}| \sin \theta_{\rm A,B}  \end{array}\right],
\end{equation}
where $\theta_{\rm B} = \theta_{\rm A}+180^{\circ}$ are the orbital phases of the individual stars. The outer offset curve, defined to be $R_{\rm A,B}$ perpendicularly outwards from Eq.~\ref{eq:CoM_curve} for all $\theta_{\rm A,B}$, is

\begin{equation}
\label{eq:outer_parallel_curve}
\left[\begin{array}{c}x_{\rm A,B, outer} \\y_{\rm A,B, outer}\end{array}\right] = \left[\begin{array}{c}\left(a_{\rm A,B}+\frac{|\cos I_{\rm bin}|R_{\rm A,B}}{\sqrt{\sin^2\theta_{\rm A,B} + \cos^2 I_{\rm bin}\cos^2\theta_{\rm A,B}}}\right)\cos \theta_{\rm A,B} \\\left(a_{\rm A,B}|\cos I_{\rm bin}|+\frac{R_{\rm A,B}}{\sqrt{\sin^2\theta_{\rm A,B} + \cos^2 I_{\rm bin}\cos^2\theta_{\rm A,B}}}\right)\sin \theta_{\rm A,B} \end{array}\right].
\end{equation}
The inner curve is described by
\begin{equation}
\label{eq:inner_parallel_curve}
\left[\begin{array}{c}x_{\rm A,B, inner} \\y_{\rm A,B, inner}\end{array}\right] = \left[\begin{array}{c}\left(a_{\rm A,B}-\frac{|\cos I_{\rm bin}|R_{\rm A,B}}{\sqrt{\sin^2\theta_{\rm A,B} + \cos^2 I_{\rm bin}\cos^2\theta_{\rm A,B}}}\right)\cos \theta_{\rm A,B} \\\left(a_{\rm A,B}|\cos I_{\rm bin}|-\frac{R_{\rm A,B}}{\sqrt{\sin^2\theta_{\rm A,B} + \cos^2 I_{\rm bin}\cos^2\theta_{\rm A,B}}}\right)\sin \theta_{\rm A,B} \end{array}\right],
\end{equation}
which differs from the outer curve only by a negative sign before the fraction, but this imposes a significant change. To better understand Eqs.~\ref{eq:outer_parallel_curve} and ~\ref{eq:inner_parallel_curve}, consider two limiting cases. When $I_{\rm bin}=0^{\circ}$ the orbit is face on and the outer and inner offset curves reduce to

\begin{equation}
\label{eq:outer_curve_0deg}
\left[\begin{array}{c}x_{\rm A,B, outer} \\y_{\rm A,B,outer}\end{array}\right]_{I_{\rm bin}=0^{\circ}} = \left[\begin{array}{c}(a_{\rm A,B}+R_{\rm A,B}) \cos \theta_{\rm A,B} \\ (a_{\rm A,B}+R_{\rm A,B}) \sin \theta_{\rm A,B}  \end{array}\right]
\end{equation}
and
\begin{equation}
\label{eq:inner_curve_0deg}
\left[\begin{array}{c}x_{\rm A,B, inner} \\y_{\rm A,B,inner}\end{array}\right]_{I_{\rm bin}=0^{\circ}} = \left[\begin{array}{c}(a_{\rm A,B}-R_{\rm A,B}) \cos \theta_{\rm A,B} \\ (a_{\rm A,B}-R_{\rm A,B}) \sin \theta_{\rm A,B}  \end{array}\right],
\end{equation}
respectively, which expectedly describes a circular ring of outer diameter $2(a_{\rm A,B}+R_{\rm A,B})$ and thickness $2R_{\rm A,B}$. In the other limit of $I_{\rm bin} = 90^{\circ}$, i.e. a perfectly edge-on eclipsing binary, the offset curves reduce to

\begin{equation}
\label{eq:outer_curve_90deg}
\left[\begin{array}{c}x_{\rm A,B, outer} \\y_{\rm A,B,outer}\end{array}\right]_{I_{\rm bin}=90^{\circ}} = \left[\begin{array}{c}a_{\rm A,B}\cos \theta_{\rm A,B} \\ R_{\rm A,B}  \end{array}\right]
\end{equation}
and
\begin{equation}
\label{eq:inner_curve_90deg}
\left[\begin{array}{c}x_{\rm A,B, inner} \\y_{\rm A,B,outer}\end{array}\right]_{I_{\rm bin}=90^{\circ}} = \left[\begin{array}{c}a_{\rm A,B}\cos \theta_{\rm A,B} \\ -R_{\rm A,B}  \end{array}\right],
\end{equation}
which expectedly describe a rectangle of length $2a_{\rm A,B}$ and height $2R_{\rm A,B}$. For $I_{\rm bin}$ at least $\sim 20^{\circ}$ away from $90^{\circ}$ the orbital extent roughly resembles an annulus. In Fig.~\ref{fig:stellar_orbit_extent} are several examples of the extent of the stellar orbit.

Since the planet is assumed to have negligible radius, comparable offset curves are not calculated. A planet is in transitability when its projected orbital ellipse (Eq.~\ref{eq:planet_orbit}) intersects the outer edge of the stellar orbit (Eq.~\ref{eq:outer_parallel_curve}). The inner offset curve plays no part in determining transitability. 

\subsection{Orbital Dynamics}

Since we are considering a small outer body, the inner binary orbit is unperturbed, so we can take all of its orbital elements to be constant, except for its orbital phase. Contrastingly, the outer planetary orbit receives significant perturbations from the binary. According to \citet{schneider94,farago10,doolin11} the orbital plane of the planet precesses around that of the binary at a constant rate, with period

\begin{equation}
\label{eq:schneider_prec_period}
T_{\rm prec} = \frac{4}{3}\left(\frac{T_{\rm p}^7}{T_{\rm bin}^4} \right)^{1/3} \frac{\left(M_{\rm A} + M_{\rm B}\right)^2}{M_{\rm A}M_{\rm B}}\frac{\left(1-e_{\rm p}^2\right)^2}{\cos \Delta I},
\end{equation}
whilst maintaining a constant mutual inclination, $\Delta I$. For any $\Delta I$ circumbinary planets can have stable orbits \citep{lohinger03,doolin11} and high-eccentricity Kozai-Lidov cycles are not applicable \citep{martin16}. Even though in this paper we consider circular binaries and planets when deriving the geometry of transitability, we can at least include planet eccentricity in calculating the precession period. Aside for a slight change to the timescale, the orbital precession behaves the same. Contrarily, if the binary is eccentric it induces some variations in $\Delta I$, in addition to changing the timescale. The precession timescale of planets around eccentric binaries was derived by \citet{farago10} and is a more complicated function whichh we do not reproduce here.  The effect on time-dependent transitability by assuming circular orbits is briefly investigated in Sect.~\ref{subsec:eccentricity}, but fully incorporating the geometry and dynamics of eccentric orbits is a future task.

For constant $\Delta I$, as the orbital plane of the planet rotates $I_{\rm p}(t)$ and $\Omega_{\rm p}(t)$ vary, whilst $a_{\rm p}$ is constant. The inclination of the planet on the plane of the sky, owing to this precession, follows a sinusoidal path

\begin{equation}
\label{eq:I_p}
I_{\rm p}(t) = \Delta I \cos\left(\frac{2\pi}{T_{\rm prec}}(t-t_0)\right) + I_{\rm bin},
\end{equation}
where $I_{\rm p,0}$ is the initial planetary inclination at time $t_0$,

\begin{equation}
\label{eq:t_0}
t_0 = -S \frac{T_{\rm prec}}{2\pi} \cos^{-1}\left(\frac{I_{\rm p,0} - I_{\rm bin}}{\Delta I} \right),
\end{equation}
 where the factor $S = -1$ if $\Omega_{\rm p,0}$ is between 0 and $180^{\circ}$ and $S=+1$ if $\Omega_{\rm p,0}$ is between 180 and $360^{\circ}$, where $\Omega_{\rm p,0}$ is the initial planetary longitude of the ascending node. By inverting Eq.~\ref{eq:Delta_Omega} and substituting in Eq.~\ref{eq:Delta_I} the time-dependent equation for $\Omega_{\rm p}(t)$ is

\begin{equation}
\label{eq:Omega_p}
\Omega_{\rm p}(t) = -\cos^{-1}\left(\frac{\cos \Delta I - \cos I_{\rm bin}\cos I_{\rm p}(t)}{\sin I_{\rm bin}\sin I_{\rm p}(t)}\right),
\end{equation}
Throughout this paper as emphasis we always state the explicit time-dependence of $I_{\rm p}(t)$ and $\Omega_{\rm p}(t)$.

\section{Derivation of time-dependent transitability}\label{sec:derivation}

Transitability occurs when the planetary ellipse (Eq.~\ref{eq:planet_orbit}) intersects the outer edge of the binary orbit (Eq.~\ref{eq:outer_parallel_curve}). An exact solution of this would require finding solutions to 

\begin{equation}
x_{\rm A,B,outer}(\theta_{\rm A,B}) = x_{\rm p}(\theta_{\rm p}) \quad {\rm and} \quad y_{\rm A,B,outer}(\theta_{\rm A,B}) =y_{\rm p}(\theta_{\rm p}).
\end{equation}
Calculating such intersections is not analytically feasible given the complicated functional form of the outer binary extent. Instead, simple geometric approach is employed which avoids explicit use of Eq.~\ref{eq:outer_parallel_curve}.

For configurations where the planet and stellar orbits do not intersect, we know that the distance between the planet and binary ellipses is minimised when the derivatives are equal:

\begin{equation}
\label{eq:equal_derivatives}
\frac{dy_{\rm A,B}}{dx_{\rm A,B}}(\theta_{\rm A,B}) = \frac{dy_{\rm p}}{dx_{\rm p}}(\theta_{\rm p}).
\end{equation}
We define this minimum distance as $d_{\rm A,B}(t)$, where the time-dependence is indicative of the planet's dynamical evolution changing its orbital orientation. A planet is inside transitability on the primary and/or secondary star at time $t$ when

 \begin{figure}
\begin{center}  
\includegraphics[width=0.5\textwidth]{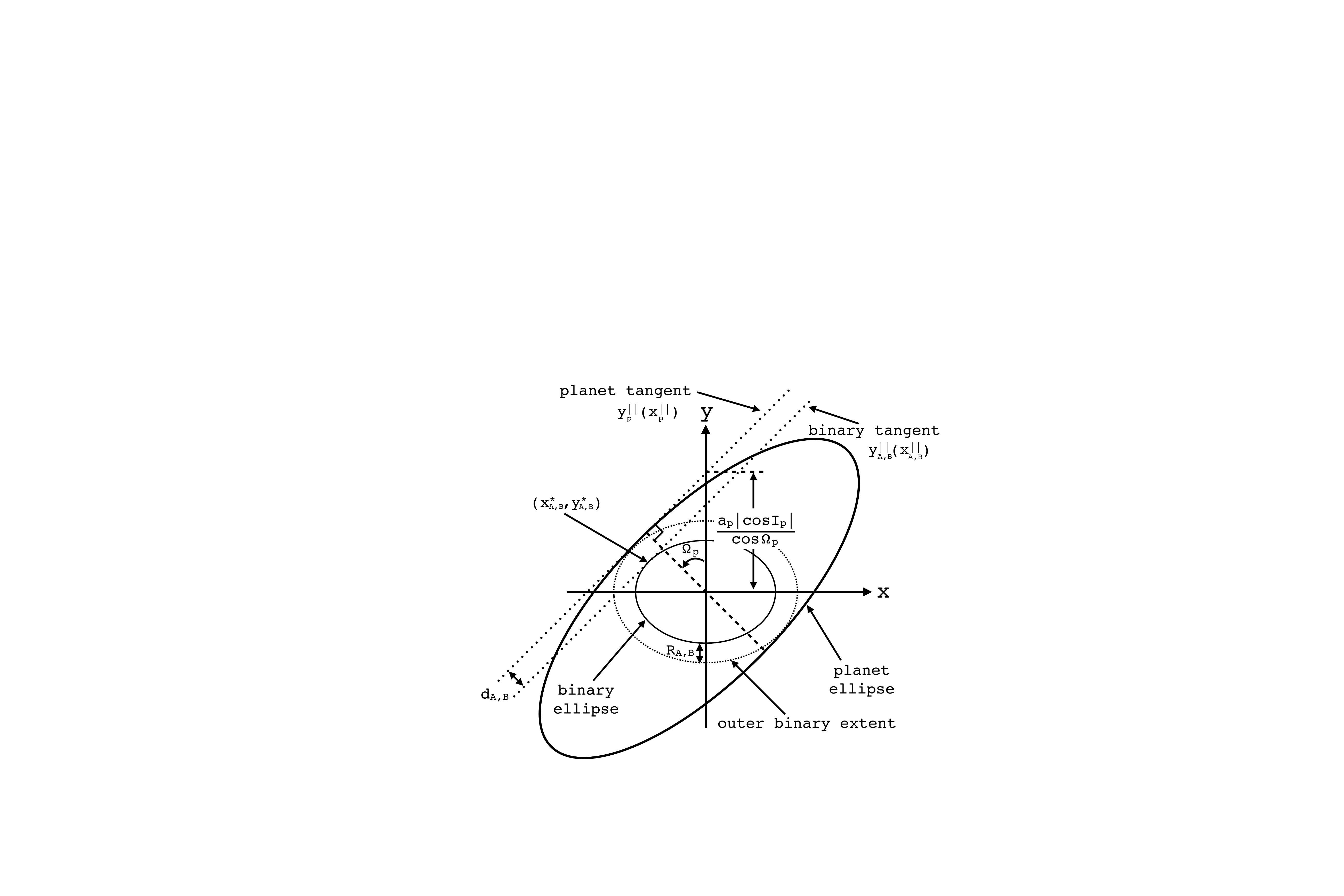}  
\caption{Illustration of the geometric definition of $d_{\rm A,B}(t)$, the minimum distance between the planet and binary ellipses for a given orientation. Only one stellar orbit is shown, where we plot both the ellipse tracing out the centre of the star (solid line) and the outer offset curve (narrow dashed line). The dotted tangent lines correspond to Eq.~\ref{eq:planet_tangent} for the planet and Eq.~\ref{eq:binary_tangent} for the binary. In this example, the planet is on the edge of transitability, and hence $d_{\rm A,B}(t) = R_{\rm A,B}$.}
\label{fig:Trig_Diagram}
\end{center}  
\end{figure} 

\begin{equation}
\label{eq:fund_transitability_def}
d_{\rm A,B}(t) < R_{\rm A,B}.
\end{equation}

To calculate $d_{\rm A,B}(t)$ we first approximate the planet orbit by a tangential line with a gradient $\tan \Omega_{\rm p}(t)$, characterised by the function
\begin{equation}
\label{eq:planet_tangent}
y_{\rm p}^{\rm ||} = \tan \Omega_{\rm p}(t)x_{\rm p}^{\rm ||}  + \frac{a_{\rm p}|\cos I_{\rm p}(t)|}{\cos \Omega_{\rm p}(t)},
\end{equation}
where the geometry of this equation is shown in Fig.~\ref{fig:Trig_Diagram}. This is a valid approach for a planet near the edge of transitability, as long as the planet orbit is sufficiently larger than the stellar orbits. Fortuitously, dynamical stability constraints dictate that $a_{\rm p} \gtrsim 3a_{\rm A,B}$ \citep{dvorak86,dvorak89,holman99}, which is sufficient to make the approximation valid.

The next step is to calculate the tangent equation to the binary ellipse with the same gradient $\tan \Omega_{\rm p}(t)$. The distance between these two parallel lines corresponds to $d_{\rm A,B}(t)$, also illustrated in Fig.~\ref{fig:Trig_Diagram}. This diagram also illustrates the origin of the last part of Eq.~\ref{eq:planet_tangent}.

To calculate the binary tangent line, first find the point on the binary orbit ellipse, $(x^*_{\rm A,B},y^*_{\rm A,B})$, where $dy_{\rm A,B}/dx_{\rm A,B} = \tan \Omega_{\rm p}(t)$. Differentiate Eq.~\ref{eq:binary_orbit} with respect to $x_{\rm A,B}$:

\begin{align}
\label{eq:differentiate}
\frac{2x_{\rm A,B}}{a_{\rm A,B}^2} +  \frac{2y_{\rm A,B}}{a_{\rm A,B}^2\cos^2 I_{\rm bin}}\frac{dy_{\rm A,B}}{dx_{\rm A,B}} = 0.
\end{align}
Evaluating Eq.~\ref{eq:differentiate} at $dy_{\rm A,B}/dx_{\rm A,B} = \tan \Omega_{\rm p}(t)$ and re-arranging yields an expression relating $y^*_{\rm A,B}$ and $x^*_{\rm A,B}$:

\begin{align}
\label{eq:y*_and_x*}
y^*_{\rm A,B} = -x^*_{\rm A,B}\frac{\cos^2 I_{\rm bin}}{\tan \Omega_{\rm p}(t)}.
\end{align}
Subsitute the expression for $y^*_{\rm A,B}$ in Eq.~\ref{eq:y*_and_x*} into the binary ellipse in Eq.~\ref{eq:binary_orbit} and solve for $x^*_{\rm A,B}$:

\begin{align}
\label{eq:expression_for_x*}
x^*_{\rm A,B} = \frac{-a_{\rm A,B}}{\sqrt{1+\frac{\cos^2I_{\rm bin}}{\tan^2\Omega_{\rm p}(t)}}},
\end{align}
where we have taken the negative root to match the diagram in Fig.~\ref{fig:Trig_Diagram}. Substitute Eq.~\ref{eq:expression_for_x*} for $x^*_{\rm A,B}$ into Eq.~\ref{eq:y*_and_x*} and solve for $y^*_{\rm A,B}$:

\begin{align}
\label{eq:expression_for_y*}
y^*_{\rm A,B}&= \frac{a_{\rm A,B}\cos^2I_{\rm bin}}{\tan\Omega_{\rm p}(t)\sqrt{1+\frac{\cos^2I_{\rm bin}}{\tan^2\Omega_{\rm p}(t)}}}.
\end{align}
Now calculate the tangent line $y_{\rm A,B}^{\rm ||}(x_{\rm A,B}^{\rm ||})$ using


\begin{align}
\begin{split}
\label{eq:binary_tangent_workings}
y_{\rm A,B}^{\rm ||} - y_{\rm A,B}^* &= \tan \Omega_{\rm p}(t)\left(x_{\rm A,B}^{\rm ||} - x_{\rm A,B}^*\right) \\
y_{\rm A,B}^{\rm ||}  - \frac{a_{\rm A,B}\cos^2I_{\rm bin}}{\tan\Omega_{\rm p}(t)\sqrt{1+\frac{\cos^2I_{\rm bin}}{\tan^2\Omega_{\rm p}(t)}}} &= \tan \Omega_{\rm p} (t)\left(x_{\rm A,B}^{\rm ||} + \frac{a_{\rm A,B}}{\sqrt{1+\frac{\cos^2I_{\rm bin}}{\tan^2\Omega_{\rm p}(t)}}}\right).
\end{split}
\end{align}
Rearrange to form the final  binary tangent equation:
\begin{align}
\label{eq:binary_tangent}
y_{\rm A,B}^{\rm ||} &= \tan\Omega_{\rm p}(t)x_{\rm A,B}^{\rm ||} + a_{\rm A,B}\sqrt{\tan^2\Omega_{\rm p}(t) + \cos^2I_{\rm bin}}.
\end{align}

The planet tangent equation (Eq.~\ref{eq:planet_tangent}) and binary tangent equation (Eq.~\ref{eq:binary_tangent}) are parallel. The shortest distance between two parallel lines $y_1 = mx_1 + c_1$ and $y_2 = mx_2 + c_2$ is

\begin{equation}
\frac{c_2 - c_1}{\sqrt{m^2+1}}.
\end{equation}
The distance between the two parallel lines in Eqs.~\ref{eq:planet_tangent} and \ref{eq:binary_tangent} is $d_{\rm A,B}(t)$, which we calculate to be
\begin{equation}
\label{eq:d_almost_final}
d_{\rm A,B}(t) = \frac{\frac{a_{\rm p}|\cos I_{\rm p}(t)|}{\cos\Omega_{\rm p}(t)} - a_{\rm A,B}\sqrt{\tan^2\Omega_{\rm p}(t) + \cos^2I_{\rm bin}}}{\sqrt{\tan^2\Omega_{\rm p}(t) + 1}},
\end{equation}
which we simplify to
\begin{equation}
\label{eq:d_final}
d_{\rm A,B}(t) = a_{\rm p}|\cos I_{\rm p}(t)| - a_{\rm A,B}\cos\Omega_{\rm p}(t)\sqrt{\tan^2\Omega_{\rm p}(t) + \cos^2 I_{\rm bin}}.
\end{equation}

It is not possible to analytically solve the inequality Eq.~\ref{eq:fund_transitability_def} with this expression for $d_{\rm A,B}(t)$ due to the multiple instances of $t$ so a simplification is needed. Transitability occurs when $I_{\rm p}$ is near $90^{\circ}$. Applying this to Eq.~\ref{eq:Omega_p}, we approximate $\Omega_{\rm p}$ in transitability as a constant value

\begin{figure}  
\begin{center}  
	\begin{subfigure}[b]{0.49\textwidth}
		\includegraphics[width=\textwidth]{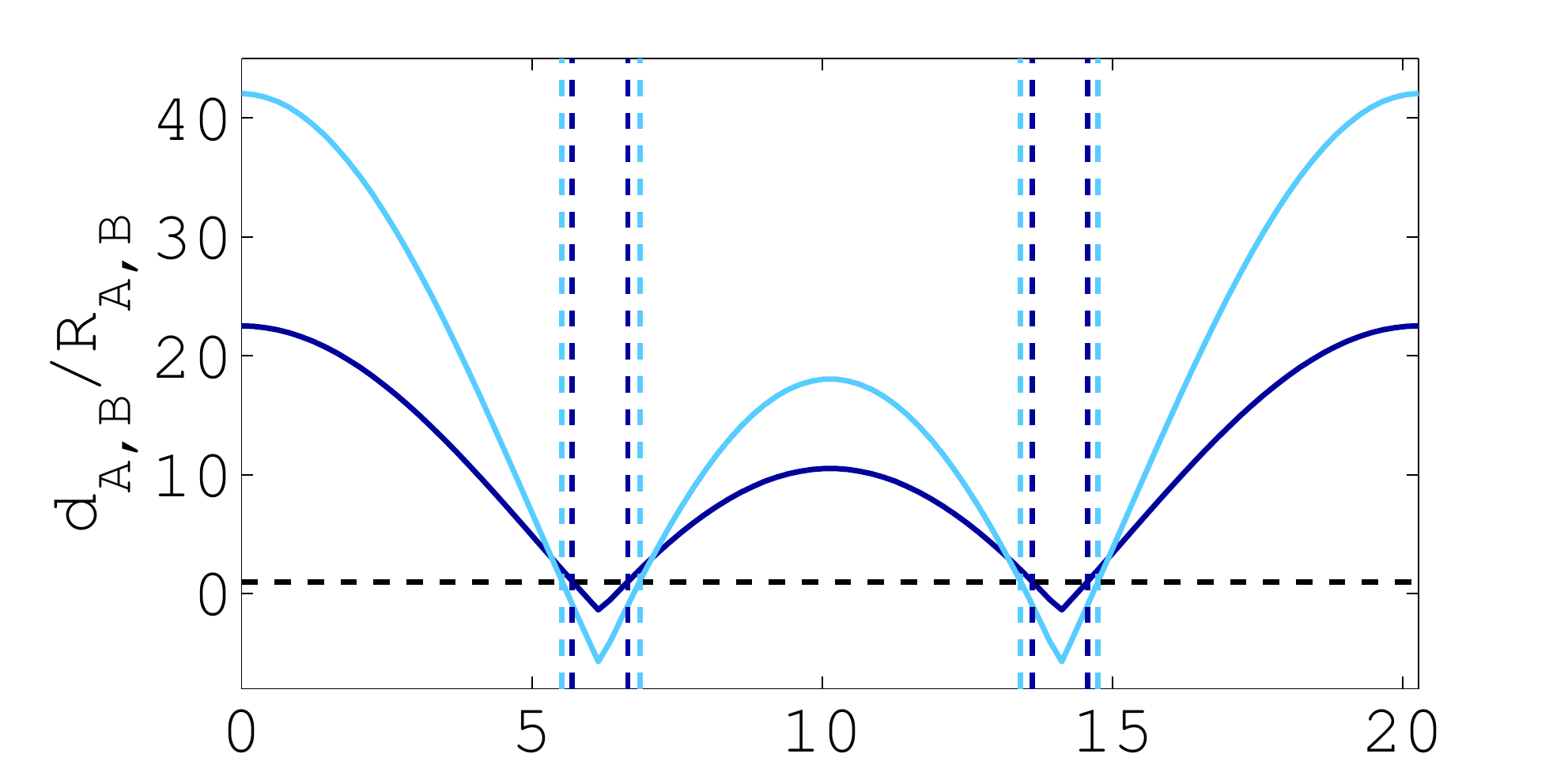}  
	\end{subfigure}
	\begin{subfigure}[b]{0.49\textwidth}
		\includegraphics[width=\textwidth]{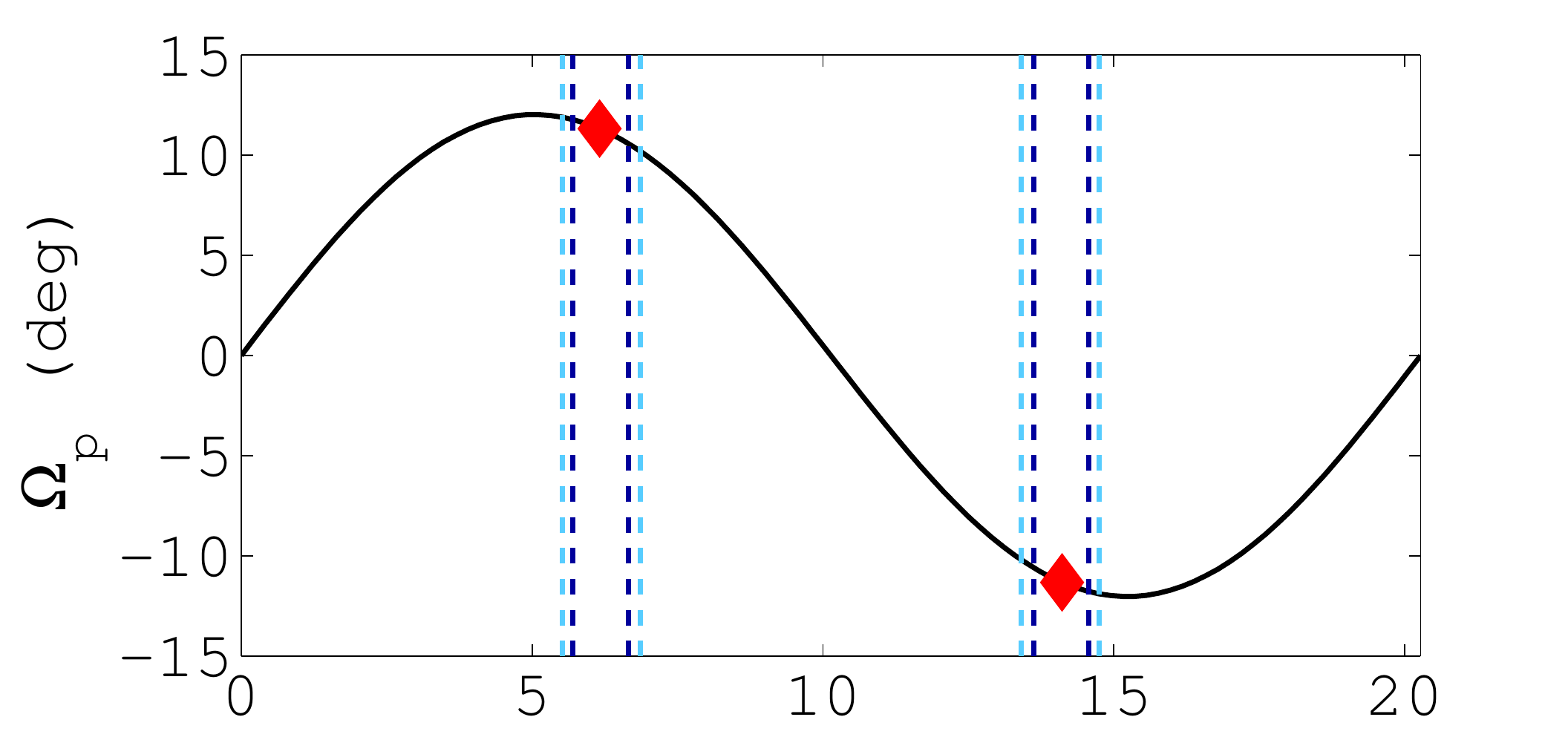}  
	\end{subfigure}
	\begin{subfigure}[b]{0.49\textwidth}
		\includegraphics[width=\textwidth]{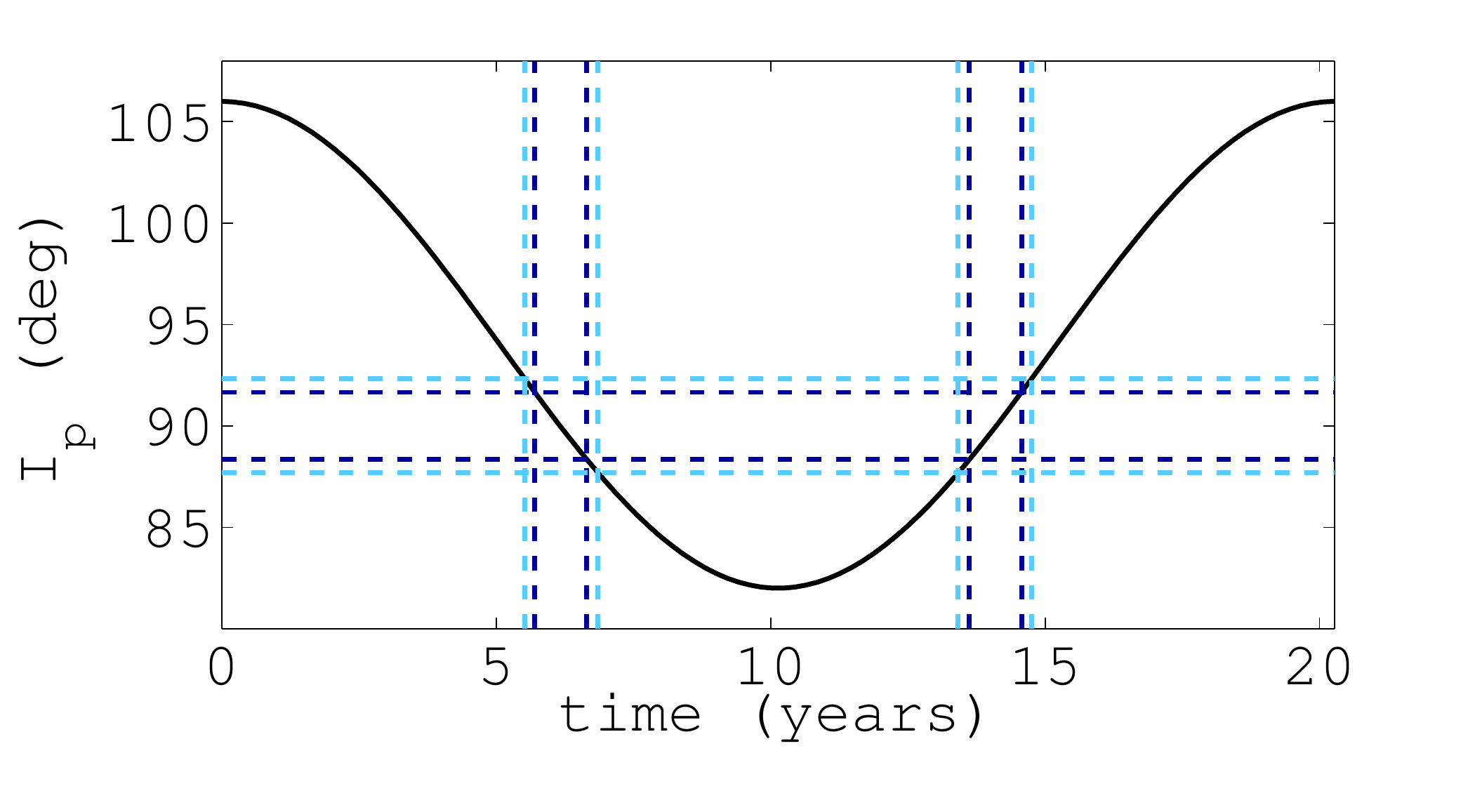}  	
	\end{subfigure}
	\caption{ Calculation of the window of transitability for an example circumbinary system: $M_{\rm A} = 1M_{\odot}$, $M_{\rm B} = 0.5M_{\odot}$, $R_{\rm A} = 1R_{\odot}$, $R_{\rm B} = 1R_{\odot}$, $a_{\rm bin}=0.1$ AU, $a_{\rm p} = 0.4$ AU, $I_{\rm bin} = 94^{\circ}$, $\Delta I = 12^{\circ}$ and $\Omega_{\rm p}(t=0)=0$. In all three figures the dark and light blue vertical lines denote the analytically-calculated regions in time when the planet is in transitability on the primary and secondary stars, respectively. In  the top image the minimum distance $d_{\rm A,B}$ (Eq.~\ref{eq:d_approx}) is plotted, scaled by the primary and secondary radii. The black horizontal dashed line is at $d_{\rm A,B}/R_{\rm A,B}=1$, and hence below this limit transitability occurs. Negative values occur when the planet and binary ellipses intersect. In  the middle image the red triangles denote the approximately constant value of $\Omega_{\rm p}$ when in transitability (Eq.~\ref{eq:Omega_p_approx}). In  the bottom image the horizontal dashed lines denote the limits of transitability in terms of $I_{\rm p}$ (Eq.~\ref{eq:Ip_transitability}).
}\label{fig:example}  
\end{center}  
\end{figure} 

\begin{equation}
\label{eq:Omega_p_approx}
\left. \Omega_{\rm p}\right|_{\rm transitability} \approx -\cos^{-1}\left(\frac{\cos \Delta I}{\sin I_{\rm bin}}\right).
\end{equation}
Insert  this approximation into Eq.~\ref{eq:d_final} to obtain
\begin{equation}
\label{eq:d_approx}
d_{\rm A,B}(t) = a_{\rm p} \left| \cos I_{\rm p}(t)\right| - a_{\rm A,B}\frac{\cos \Delta I}{\sin I_{\rm bin}}\sqrt{\tan^2\left[\cos^{-1}\left(\frac{\cos \Delta I}{\sin I_{\rm bin}}\right) \right] + \cos^2I_{\rm bin}}.
\end{equation}
 With only one instance of $t$ remaining in Eq.~\ref{eq:d_approx} the limits of transitability at $d_{\rm A,B}=R_{\rm A,B}$, in terms of the planet sky inclination,  can be solved for:

\begin{multline}
\label{eq:Ip_transitability}
\left. I_{\rm p}\right|_{\rm transitability} =  \pm \cos^{-1}\Bigg[\frac{R_{\rm A,B}}{a_{\rm p}} + \frac{a_{\rm A,B}}{a_{\rm p}}\frac{\cos \Delta I}{\sin I_{\rm bin}} \Bigg. \\ 
\Bigg. \times \sqrt{\tan^2\left(\cos^{-1}\left[\frac{\cos \Delta I}{\sin I_{\rm bin}} \right] \right) + \cos^2 I_{\rm bin}}  \quad \Bigg].
\end{multline}
The corresponding times that transitability is entered and exited simply come from solving Eq.~\ref{eq:I_p} for $t$ using Eq.~\ref{eq:Ip_transitability}. Depending on the parameters, there may be zero, one or two regions of transitability within a precession period, and hence zero, two or four times $t$ to solve for. In Fig.~\ref{fig:example} we provide an example of the evolution of $d_{\rm A,B}/R_{\rm A,B}$, $\Omega_{\rm p}$ and $I_{\rm p}$ over a precession period for a circumbinary planet that goes in and out of transitability twice. The secondary star has a slightly greater window of transitability in this example, because even though it has a smaller radius, it sweeps out a larger area on the sky since $a_{\rm B} = 2a_{\rm A}$ here. Secondary star transitability is usually longer except for eclipsing binaries.

 In the limit of $a_{\rm A,B}\rightarrow 0$, i.e. when the binary is compacted to a single object, the limits of transitability in Eq.~\ref{eq:Ip_transitability} reduce to

\begin{equation}
\label{eq:d_approx}
\lim_{a_{\rm A,B}\rightarrow 0}\left. I_{\rm p}\right|_{\rm transitability} = \pm\cos^{-1}\left(\frac{R_{\rm A,B}}{a_{\rm p}} \right),
\end{equation}
which are the inclination limits for transits of a single star, as expected.

\section{Analysis}\label{sec:analysis}

\subsection{Will the planet ever reach transitability?}\label{subsec:MT15_connection}

In this section we reproduce the result of \citet{martin15} for time-independent transitability. To know whether or not transitability will occur at some unspecified point in the planet's orbital evolution, calculate $d_{\rm A,B}(t)$ at the extrema of $I_{\rm p}(t)$, which are simply $I_{\rm bin} \pm \Delta I$. The corresponding values of $\Omega_{\rm p}$ according to Eq.~\ref{eq:Omega_p} are
\begin{equation}
\begin{aligned} 
&\left.\Omega_{\rm p}\right|_{\rm extrema} =  -\cos^{-1}\left(\frac{\cos \Delta I - \cos I_{\rm bin}\cos (I_{\rm bin} \pm \Delta I)}{\sin I_{\rm bin}\sin (I_{\rm bin} \pm \Delta I)}\right) \\ 
&=  -\cos^{-1}\left( \frac{\cos \Delta I - \frac{1}{2}\left[ \cos(I_{\rm bin}-( I_{\rm bin} \pm \Delta I)) + \cos(I_{\rm bin} + (I_{\rm bin} \pm \Delta I) \right]}{\frac{1}{2}\left[ \cos(I_{\rm bin}-(I_{\rm bin} \pm \Delta I)) - \cos(I_{\rm bin} + (I_{\rm bin} \pm \Delta I)) \right]}\right) \\
&=  -\cos^{-1}\left( \frac{\cos \Delta I - \frac{1}{2}\left[ \cos(\mp \Delta I) + \cos(2I_{\rm bin} \pm \Delta I) \right]}{ \frac{1}{2}\left[ \cos(\mp\Delta I) - \cos(2I_{\rm bin} \pm \Delta I) \right]}\right) \\
&=  -\cos^{-1}( 1) \\
&= 0,
\end{aligned}
\end{equation}
where we have used a prosthaphaeresis trigonometric identity between the first and second lines. The minimum value of $d_{\rm A,B}(t)$, according to Eq.~\ref{eq:d_final} with $\left.\Omega_{\rm p}\right|_{\rm extrema}=0$ and $I_{\rm p} = I_{\rm bin} - \Delta I$, is


\begin{equation}
\label{eq:d_min}
d_{\rm min} = a_{\rm p}\sin \left| \frac{\pi}{2}-I_{\rm bin}+\Delta I\right| - a_{\rm A,B}\sin \left| \frac{\pi}{2}-I_{\rm bin}\right|,
\end{equation}
where to match the notation of \citet{martin15} we use use $|\cos I_{\rm bin}| = \sin|\pi/2-I_{\rm bin}|$, which is valid for $I_{\rm bin} \in [0^{\circ},180^{\circ}]$.
For transitability to occur at some point requires $d_{\rm min}<R_{\rm A,B}$. Inserting this condition into Eq.~\ref{eq:d_min}\footnote{Note: we are not using the simplified version of $d_{\rm A,B}(t)$ in Eq.~\ref{eq:d_approx}.} yields
\begin{equation}
\label{eq:transitability_criterion_1}
R_{\rm A,B} > a_{\rm p}\sin \left| \frac{\pi}{2}-I_{\rm bin}+\Delta I\right| - a_{\rm A,B}\sin \left| \frac{\pi}{2}-I_{\rm bin}\right|,
\end{equation}
which we re-arrange to form
\begin{equation}
\label{eq:transitability_criterion_2}
R_{\rm A,B} + a_{\rm A,B}\sin \left| \frac{\pi}{2}-I_{\rm bin}\right|> a_{\rm p}\sin \left| \frac{\pi}{2}-I_{\rm bin}+\Delta I\right|,
\end{equation}
\begin{equation}
\label{eq:transitability_criterion_3}
\sin^{-1}\left(\frac{R_{\rm A,B} + a_{\rm A,B}\sin \left| \frac{\pi}{2}-I_{\rm bin}\right|}{a_{\rm p}}\right)> \left| \frac{\pi}{2}-I_{\rm bin}+\Delta I\right|,
\end{equation}
\begin{equation}
\label{eq:transitability_criterion_3}
\sin^{-1}\left({\frac{R_{\rm A,B}}{a_{\rm p}} + \frac{a_{\rm A,B}}{a_{\rm p}}\sin \left| \frac{\pi}{2}-I_{\rm bin}\right|}\right) - \left|\frac{\pi}{2}-I_{\rm bin}\right|> \Delta I,
\end{equation}
\begin{equation}
\label{eq:criterion_transitability_time-indep}
\Delta I > \left|\frac{\pi}{2}-I_{\rm bin}\right| - \sin^{-1}\left({\frac{a_{\rm A,B}}{a_{\rm p}}\sin \left| \frac{\pi}{2}-I_{\rm bin}\right| + \frac{R_{\rm A,B}}{a_{\rm p}}}\right),
\end{equation}
which recovers the {\it time-independent} transitability criterion derived in \citet{martin15}\footnote{Equations 18 and 19 in that paper, which use $\Delta I = |I_{\rm p}-I_{\rm bin}|$ when $\Omega_{\rm p}=0$.}.

\begin{figure}  
\begin{center}  
\includegraphics[width=0.5\textwidth]{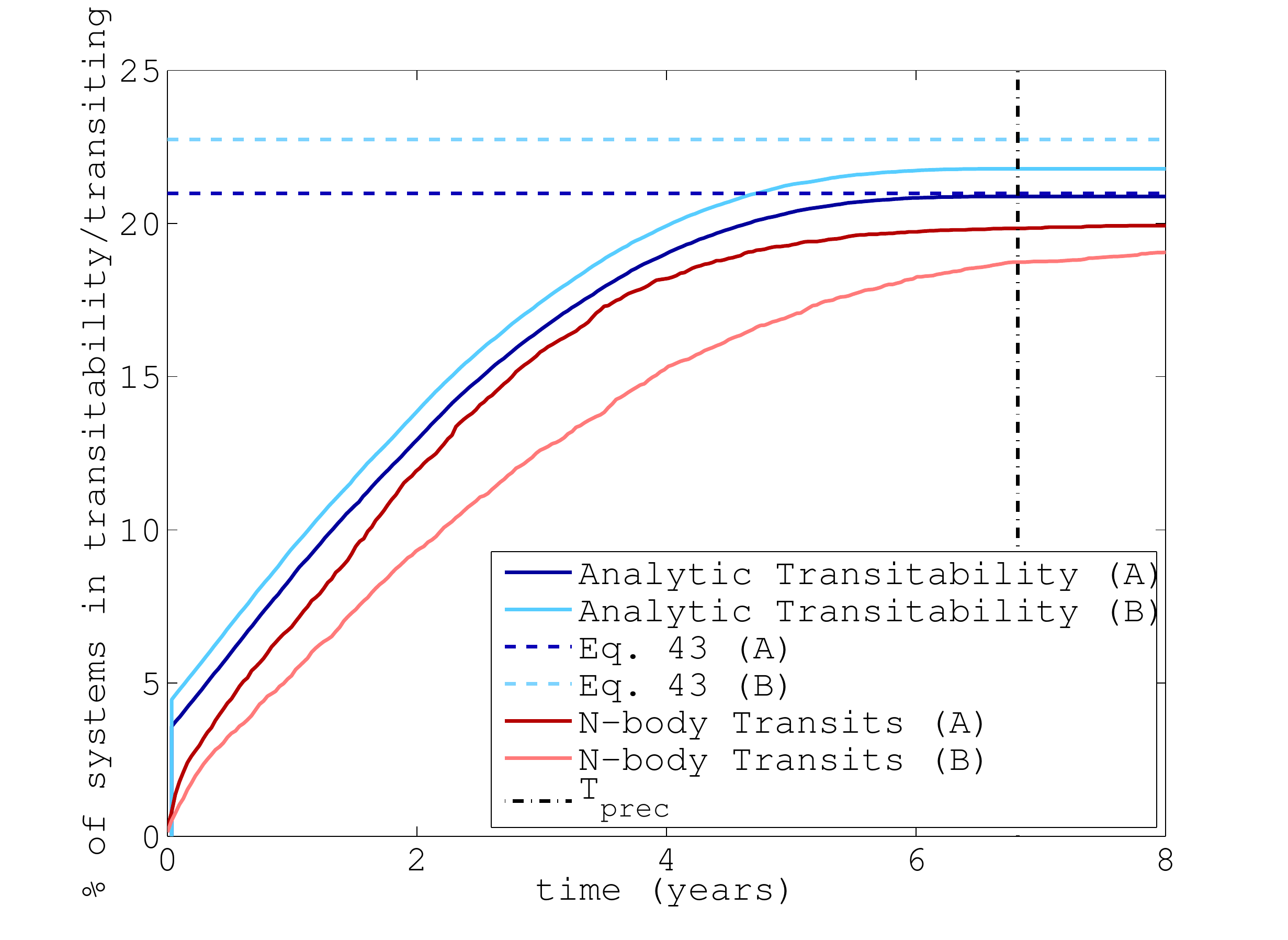}  
\caption{Transitability and transit percentages over time for an example circumbinary system with $M_{\rm A} = 1M_{\odot}$, $M_{\rm B} = 0.5 M_{\odot}$, $R_{\rm A} = 1R_{\odot}$, $R_{\rm B} = 0.5 R_{\odot}$, $T_{\rm bin} = 7$ d, $T_{\rm p} = 40$ d and $\Delta I = 10^{\circ}$. In dark and light blue solid lines we plot the {\it time-dependent} probability of transitability $P_{\rm A,B}(t)$ as a percentage for the primary and secondary stars, respectively. In dark and light red solid lines we plot the percentage of systems found to be actually transiting at time $t$ using an N-body code. The horizontal dashed blue lines at the top indicate the {\it time-independent} probability of transitability from Eq.~\ref{eq:MT15}. Finally, the black vertically dot-dashed indicates $T_{\rm prec} = 6.82$ yr. 
}
\label{fig:Test_vs_Nbody}
\end{center}  
\end{figure}

\begin{figure*}  
\begin{center}  
	\begin{subfigure}[b]{0.49\textwidth}
		\caption{$a_{\rm p} = 0.3553$ AU, varied $\Delta I$ and $I_{\rm bin}$}
		\includegraphics[width=\textwidth]{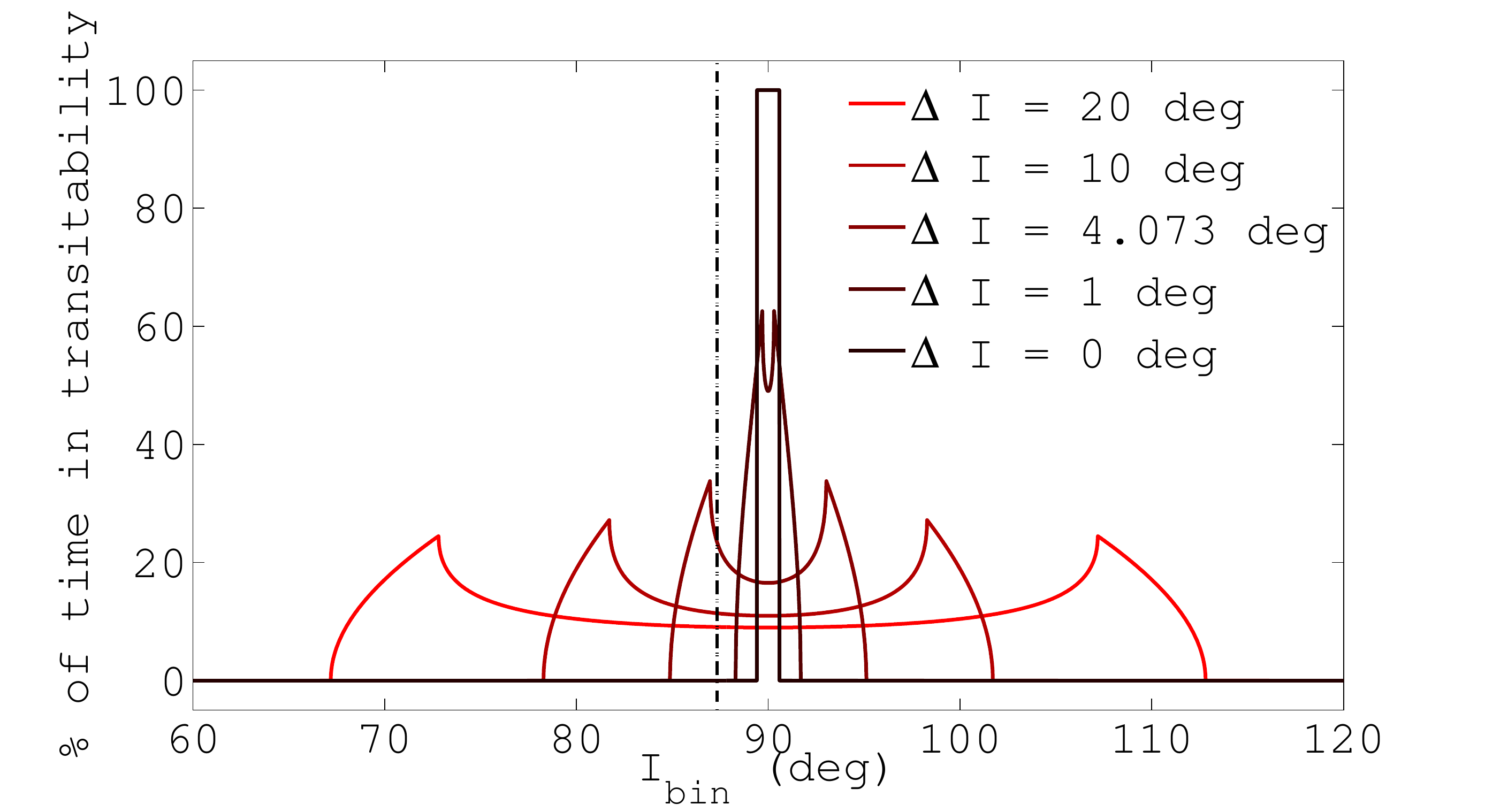}  
		\label{fig:Percentage_Test_a}
	\end{subfigure}
	\begin{subfigure}[b]{0.49\textwidth}
		\caption{$\Delta I = 4.073^{\circ}$, varied $a_{\rm p}$ and $I_{\rm bin}$}
		\includegraphics[width=\textwidth]{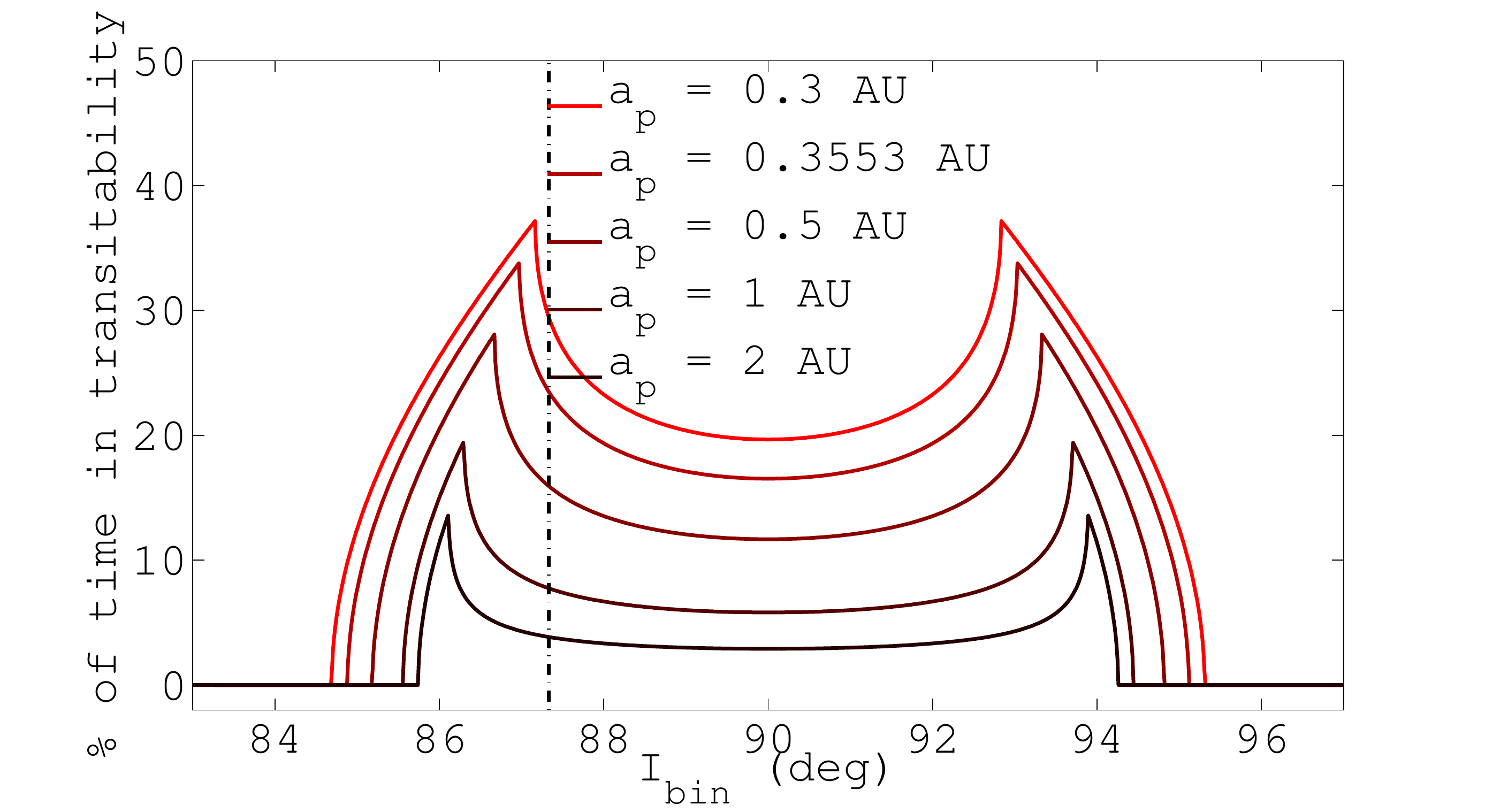}  
		\label{fig:Percentage_Test_b}
	\end{subfigure}

	\caption{The analytically-calculated percentage of time that a planet spends in transitability on the primary star as a function of $I_{\rm bin}$. The orbital parameters are taken from Kepler-413 \citep{kostov14}: $M_{\rm A} = 0.820 M_{\odot}$, $M_{\rm B}=0.542 M_{\odot}$, $R_{\rm A} = 0.78 R_{\odot}$, $a_{\rm bin} = 0.10148$ AU, $a_{\rm p} = 0.3553$ AU, $\Delta I =4.073^{\circ}$. All values of $I_{\rm bin}$ outside of the plotted range have 0\% transitability. The true binary inclination $I_{\rm bin}=87.33^{\circ}$ is demarcated by a black vertical dot-dashed line. In a) we vary the mutual inclination from its nominal value, with the red to black colour gradient denoting a decreasing $\Delta I$. In b) we instead value the planet semi-major axis, with the red to black colour gradient denoting an in creasing $a_{\rm p}$.}\label{fig:Percentage_Test}  
\end{center}  
\end{figure*}

\subsection{Time-dependent probability of transitability}\label{subsec:probability}

In \citet{martin15} we calculated that the probability of a circumbinary planet exhibiting transitability at some unspecified point in time is
\begin{equation}
\label{eq:MT15}
P_{\rm A,B} = \sin \left( \Delta I + \frac{a_{\rm A,B}\sin\Delta I + R_{\rm A,B}}{a_{\rm p}-a_{\rm A,B}\cos\Delta I} \right),
\end{equation}
where this equation assumes $\cos I_{\rm bin}$ is uniformly distributed, and hence it covers both eclipsing and non-eclipsing binaries\footnote{In the published version of \citet{martin15} this equation (Eq. 24 in that paper) contains a typo where the $-$ sign in the denominator is incorrectly a $+$ sign. Similarly, Eq. 22 of that paper has $+$ sign that should be a $-$ sign, and Eq. 23 has a $-$ sign that should be a $+$ sign. Those errors were purely typographical and the results presented throughout that paper were done using the correct formulae. Furthermore, the typos have been fixed in the arXiv version of the paper. We are sorry for the errors and any inconvenience caused.}. We may improve upon this by calculating $P_{\rm A,B}(t)$ using the new time-dependent criteria for transitability.

To calculate $P_{\rm A,B}(t)$ we create a uniform distribution of $\cos I_{\rm bin}$ and for each value of $I_{\rm bin}$ we choose a random $t_0$ between $0$ and $T_{\rm prec}$. With these two values and the other set system parameters we can analytically solve $I_{\rm p}(t) = \left. I_{\rm p} \right|_{\rm transitability}$ for $t$ and find the time the system first enters transitability. The probability $P_{\rm A,B}(t)$ is by definition the fraction of systems which have already entered transitability by the time $t$. At $t = T_{\rm prec}$ all systems that will ever enter transitability will have already done so, at which point $P_{\rm A,B}(t)$ should reach the value calculated in Eq.~\ref{eq:MT15}.

In Fig.~\ref{fig:Test_vs_Nbody} we show an example calculation for a circumbinary system with $M_{\rm A} = 1M_{\odot}$, $M_{\rm B} = 0.5 M_{\odot}$, $R_{\rm A} = 1R_{\odot}$, $R_{\rm B} = 0.5 R_{\odot}$, $T_{\rm bin} = 7$ d, $T_{\rm p} = 40$ d and $\Delta I = 10^{\circ}$. This is the same test as was done in Fig.~11 of \citet{martin15}. For both primary and secondary stars we plot $P_{\rm A,B}(t)$, transit probabilities calculated using an N-body code and $P_{\rm A,B}$ coming from Eq.~\ref{eq:MT15}.

As expected, the curves of $P_{\rm A,B}(t)$ are higher than the N-body transit probabilities. This is because transitability is not 100\% efficient at producing transits. The analytic and N-body curves are reasonably close for the primary star, implying a high efficiency of transitability. On the other hand, transitability is significantly less efficient on the secondary star, which is expected, since the secondary star is both physically smaller and it sweeps out a wider region of the sky, so it is easier for a planet to miss transits. The tricky process of calculating this analytic efficiency of transitability is to be done in the third and final paper of this series. We note that the $P_{\rm A,B}(t)$ curves do not start at zero at $t=0$ because some systems begin in transitability.

 There is one problem evident in Fig.~\ref{fig:Test_vs_Nbody}: the analytically calculated $P_{\rm A,B}(t)$ at $t=T_{\rm prec}$ does not quite reach the values calculated in Eq.~\ref{eq:MT15}. For the primary star (dark blue) this is barely noticeable but this small discrepancy is readily apparent for the secondary star (light blue). This small error is the result of the Eq.~\ref{eq:Omega_p_approx} approximation of constant $\Omega_{\rm p}$ during transitability, which was necessary to analytically derive the inclination limits for transitability in Eq.~\ref{eq:Ip_transitability}. A way to avoid this error would be to test for transitability by solving $d(t)<R_{\rm A,B}$ directly using Eq.~\ref{eq:d_final} without the approximation in Eq.~\ref{eq:Omega_p_approx}. This would require a numerical algorithm, but would nevertheless be much faster still than a large suite of N-body simulations\footnote{For example, the N-body curves in Fig.~\ref{fig:Test_vs_Nbody} were calculated using a suite of 10,000 randomised circumbinary systems and required several hours to numerically integrate the orbits and calculate transit times.}.

\subsection{Percentage of time spent in transitability}\label{subsec:percentage}
 \begin{figure}  
\begin{center}  
	\begin{subfigure}[b]{0.49\textwidth}
		\caption{Kepler 16}
		\includegraphics[width=\textwidth]{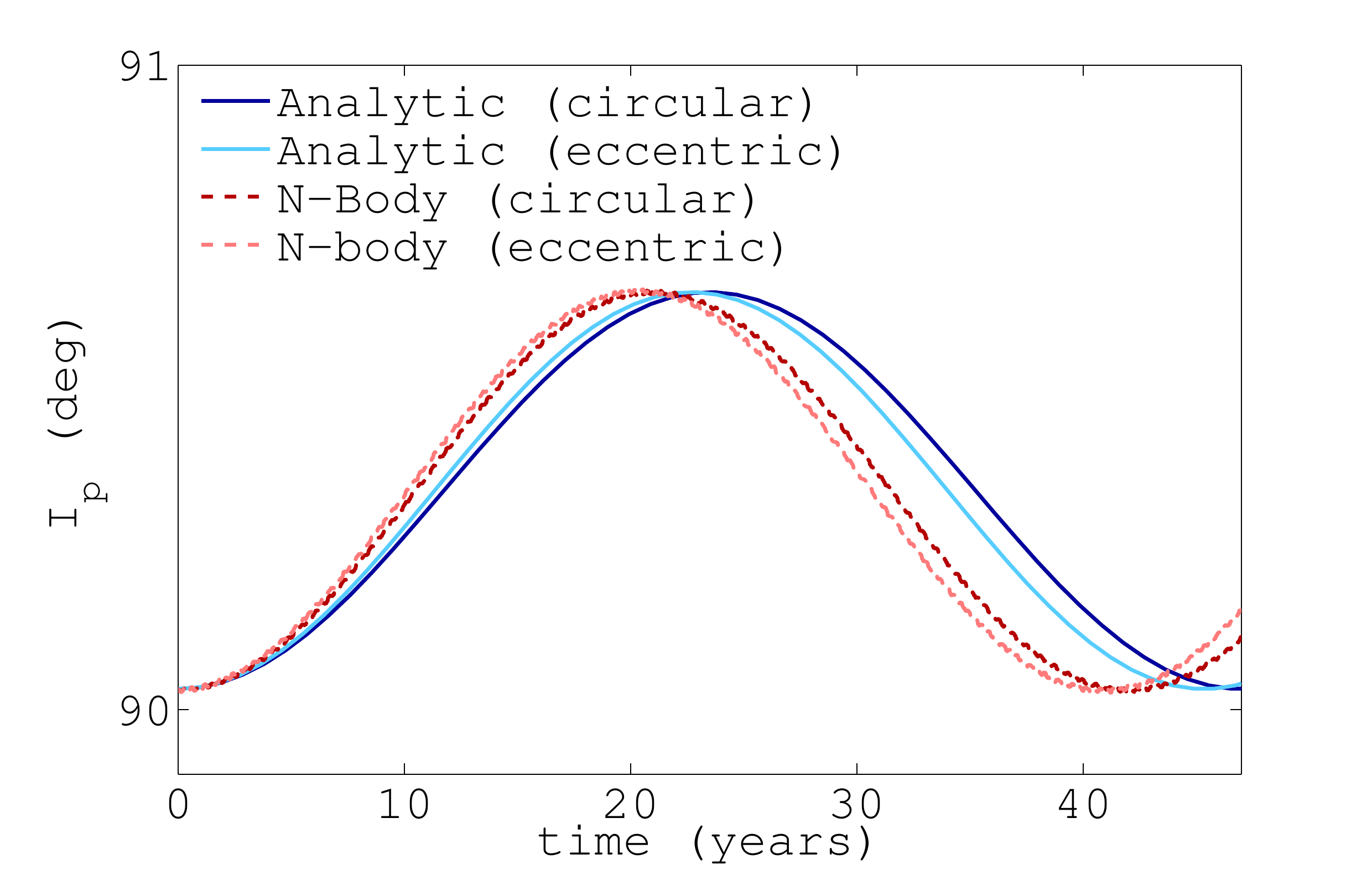}  
	\end{subfigure}
	\begin{subfigure}[b]{0.49\textwidth}
		\caption{Kepler 34}
		\includegraphics[width=\textwidth]{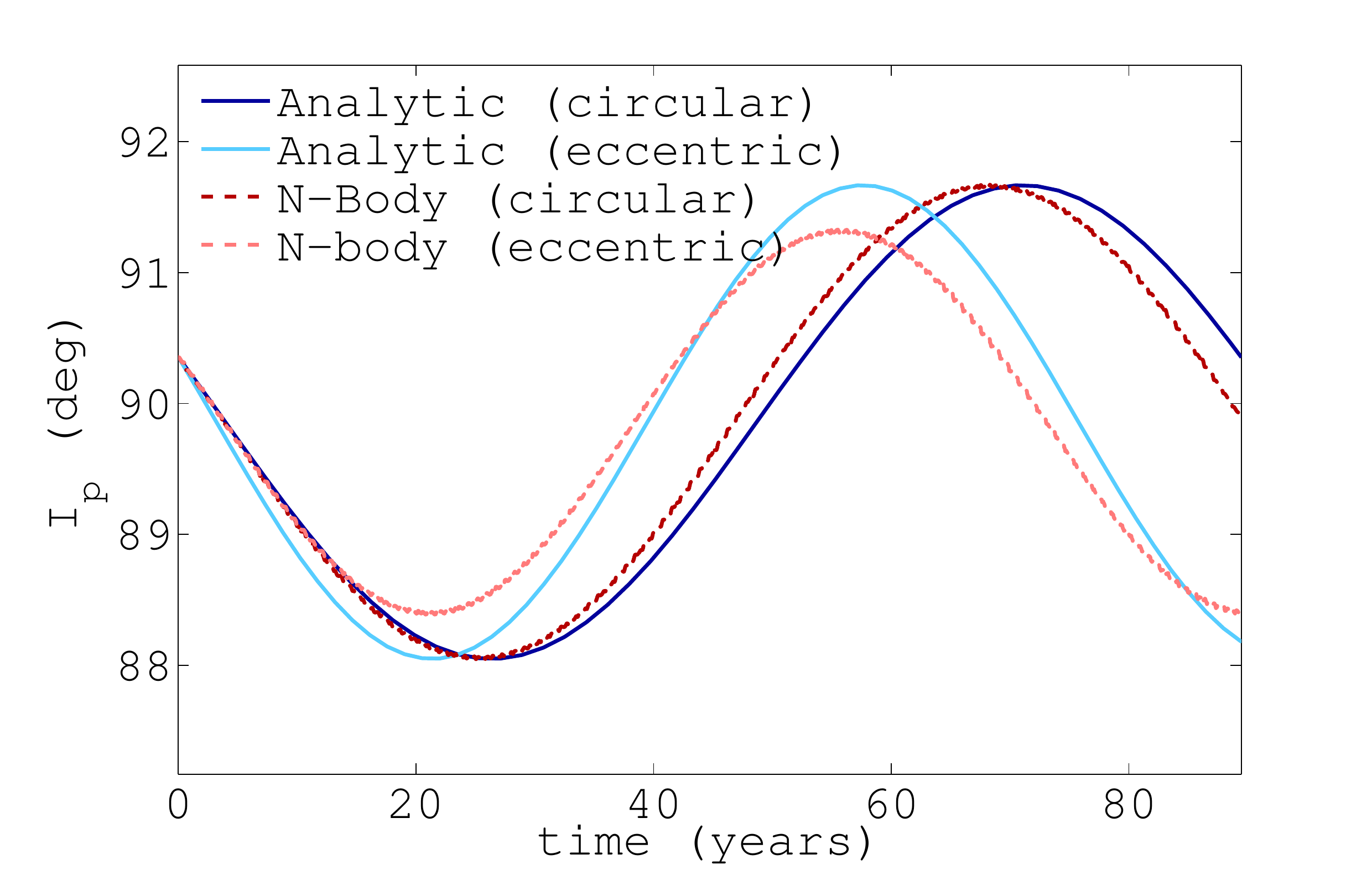}  
	\end{subfigure}
		\caption{Time evolution of $I_{\rm p}$ for Kepler-16 (a) and Kepler-34 (b), calculated in four different ways: 1) dark blue, solid line: analytically using Eq.~\ref{eq:schneider_prec_period} assuming $e_{\rm bin}=0$; 2) light blue, solid line: analytically using equations in \citet{farago10}  with $e_{\rm bin}=0.159$; 3) dark red, dashed line: numerically using N-body simulations and $e_{\rm bin}=0$ and 4) light red, dashed line: numerically using N-body simulations with $e_{\rm bin}=0.521$. In all cases $e_{\rm p}$ is set to the true value, not zero.
		}\label{fig:prec_period_test}  
\end{center}  
\end{figure} 

Whether or not a planet spends a large amount of its time in transitability or just has fleeting appearances has consequences on its detectability. As an example, the most misaligned circumbinary planet known to date is Kepler-413 with $\Delta I = 4.02^{\circ}$ \citep{kostov14}. It is also one of the tightest systems found, with $a_{\rm bin}=0.1$ AU and $a_{\rm p}=0.36$ AU, yielding a relatively short precession period of 11.1 yr. Near the beginning of the {\it Kepler} mission it transited three times, roughly 63 days apart, before disappearing for 838 days as the planet precessed out of transitability. Such a system could be easily mistaken as a transient false positive, but luckily it returned for five more transits within the original {\it Kepler} mission. In fact, Kepler-413 only spends 23.5\% and 24.0\% of its time in transitability on the primary and secondary stars, respectively.

In Fig.~\ref{fig:Percentage_Test_a} we plot the percentage of time spent in transitability as a function of $I_{\rm bin}$, for a circumbinary system with the other parameters matching Kepler-413 (see Table~\ref{tab:KeplerDiscoveries}), but with five different values of $\Delta I$. For clarity, only primary transitability is shown. As calculated in \citet{martin15}, the amount of systems exhibiting transitability (i.e. the range of $I_{\rm bin}$) {\it increases} as $\Delta I$ increases. However, the new result is that for systems exhibiting transitability, the percentage of time spent in transitability generally {\it decreases} as $\Delta I$ increases. Only for $\Delta I$ very close to $0^{\circ}$ is transitability permanent, but this only applies for $I_{\rm bin}$ very close to $90^{\circ}$. For $I_{\rm bin}= 87.33^{\circ}$, corresponding to the actual Kepler-413 system and demarcated by a black vertical dot-dashed line, a few degrees of mutual inclination is needed for transits to be possible; highly coplanar planets would never have been discovered.

In Fig.~\ref{fig:Percentage_Test_b}  we instead keep $\Delta I$ at its true value of $4.073^{\circ}$ and vary $a_{\rm p}$. For $a_{\rm p}$ between 0.3 and 2 AU there is not a significant difference in the range of $I_{\rm bin}$ centred on $90^{\circ}$ that allows transitability. This is in line with the weak period dependence found by \citet{martin15}. However, the new result is that the percentage of time in transitability is reduced as the planet is moved farther out. Distant planets may still transit but their photometric appearances are ephemeral. Furthermore, since $T_{\rm prec} \propto T_{\rm p}^{7/3}$, the time between these fleeting transit opportunities is appreciable. The one advantage of long period planets is that the efficiency of transitability should be higher for the same period binary. This is because longer-period planets move at slower speeds ($v\approx2\pi a/T$), so as the planet passes the binary the binary may cover more of its orbit and is hence less likely to be missed. The effect of this is to be quantified in the third and final paper of this series.

\subsection{Accuracy of the precession period}\label{subsec:precession_accuracy}

Because transitability is a sensitive function of $I_{\rm p}(t)$, the time-dependence of transitability is intrinsically linked to the precession period. Therefore, our ability to analytically predict windows of transitability is reliant upon the accuracy of $T_{\rm prec}$. An in-depth numerical critique of the analytic formulae from \citet{farago10} was done by \citet{doolin11}, so here we just show two examples.

 First, shown in Fig.~\ref{fig:prec_period_test}a is the evolution of $I_{\rm p}(t)$ for Kepler-16, calculated both  analytically using \citet{farago10} (blue solid curves) and from N-body simulations (red dashed curves). Results are shown for both the true binary eccentricity $e_{\rm bin}=0.16$ (lighter coloured curves) and for $e_{\rm bin}=0$ (darker coloured curves). The planet eccentricity is set to the true value in all cases, but is a negligibly small $e_{\rm p}=0.0069$. All orbital parameters are listed in Table~\ref{tab:KeplerDiscoveries}. Eccentric precession periods are, as predicted, shorter than for circular orbits. This is a small difference compared to the discrepancy of roughly 10\% error between the longer analytic periods and shorter N-body periods.

Repeating the task for Kepler-34, which has the largest eccentricity of any of the known planet hosts at $e_{\rm bin}=0.521$, we see in Fig.~\ref{fig:prec_period_test}b that there is a much more stark shortening of the precession period for the eccentric orbit. Furthermore, in this case we see a much better match of precession periods between the analytic and N-body solutions. There is, however, a small difference in the amplitude of the variation of $I_{\rm p}(t)$ between analytic and N-body curves in the eccentric case. This is because in the N-body curve $\Delta I$ varies by an amplitude of $\sim 1^{\circ}$, whereas Eq.~\ref{eq:I_p} assumes constancy. 

It is speculated that that discrepancies in the precession period may arise from the formula being calculated using a quadrupole expansion of the Hamiltonian, and higher-order effects may account for the error.  This is consistent with the \citet{farago10} quadrupole precession period working better for Kepler-34 than for Kepler-16, as the former has nearly equal mass binaries and hence the octupole perturbation on the planet is minimal. Fully quantifying this is left for future investigation.

\subsection{The effects of eccentric planets and binaries}\label{subsec:eccentricity}

 The probability of a planet transiting a single star is often simply quoted as $P_{\rm single} = R_{\rm star}/a_{\rm p}$, however when eccentricity is included \citet{barnes07} modified the equation to

\begin{equation}
\label{eq:single_star_prob_marginalised}
P_{\rm single} = \frac{R_{\rm star}}{a_{\rm p}}\frac{1}{1-e_{\rm p}^2},
\end{equation}
which has been marginalised over all possible values of $\omega_{\rm p}$. Eccentricity gives a boost to transit probabilities around single stars.

If a circumbinary system has eccentric binary and/or planetary orbits, there are three effects on the transit probability. First, like in the single star case the geometry is complexified by the addition of extra orbital elements: $e_{\rm bin}$, $\omega_{\rm bin}$, $e_{\rm p}$ and $\omega_{\rm p}$. The projected stellar orbits (e.g. Fig.~\ref{fig:stellar_orbit_extent}) are no longer vertically symmetric and the planet orbit is no longer rotationally symmetric. 

The second effect is that increased eccentricity in either the binary or planet orbit pushes the stability limit farther out \citep{holman99,mardling01}. For example, a planet with $a_{\rm p} = 3a_{\rm bin}$ is very close to the stability limit for circular orbits. However, if the planet instead has a moderate eccentricity, say more than 0.2, then its transit probability may be increased but the orbit is likely unstable. To achieve stability $a_{\rm p}$ would have to be increased, likely offsetting any gain in the transit probability. 

Finally, eccentricity affects the orbital dynamics. If $e_{\rm bin}>0$ then the assumption of constant $\Delta I$ is no longer valid and the planetary orbit precesses at a variable rate \citep{farago10,doolin11}. For $\Delta I \lesssim 45^{\circ}$ the precession is still prograde, but above this islands of libration appear and hence complicate matters further. If the planet is eccentric then $\Delta I$ remains constant and there are no islands of libration, but the constant precession period is decreased by a factor $(1-e^2)^2$, as accounted for in Eq.~\ref{eq:schneider_prec_period}. If the planet is eccentric then in addition to precession of $\Omega_{\rm p}$ there will be an apsidal advance of $\omega_{\rm p}$ at a nearly equal rate but in opposite directions \citep{lee07}. 

A full incorporation of the geometry and dynamics of eccentricity  systems is a future task. Fortunately, it was already shown in \citet{martin15} that assuming circular orbits is generally reasonable for predicting if transitability occurs. In that paper N-body simulations of 10,000 circumbinary systems were integrated over an entire precession period to check if the planet and binary orbits ever overlapped. Orbital parameters were randomised within ranges roughly corresponding to the known systems. For eccentricity $e_{\rm bin}$ and $e_{\rm p}$ were independently randomised between 0 and 0.5. Predicting whether or not transitability occurred was shown to be accurate more than 98\% of the time. Furthermore, it was found that slightly more systems entered transitability than expected when eccentricity was included, similar to the result for single stars in Eq.~\ref{eq:single_star_prob_marginalised}.

\begin{figure*}  
\begin{center}  
\includegraphics[width=1\textwidth]{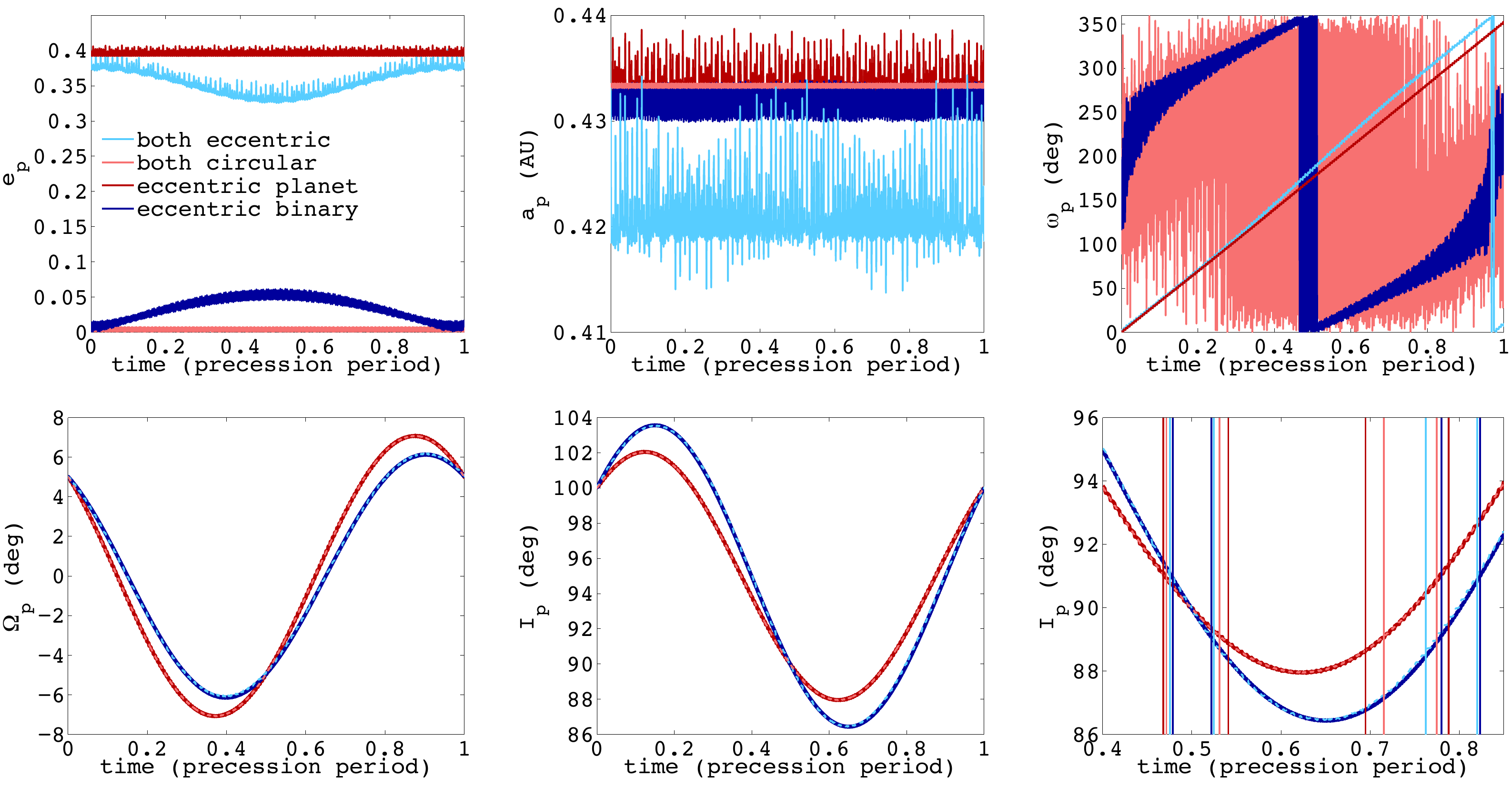}  
\caption{Variation of the orbital elements of a circumbinary planet over time according to four N-body simulations. Common to all four simulations are the starting conditions: $M_{\rm A}=1M_{\odot}$, $M_{\rm B}=0.5M_{\odot}$, $P_{\rm bin}=5$ d, $P_{\rm p} = 85$ d, $a_{\rm bin} = 0.065$ AU, $a_{\rm p} = 0.43$ AU, $I_{\rm bin} = 95^{\circ}$, $I_{\rm p}=100^{\circ}$, $\Omega_{\rm bin}=0^{\circ}$, $\Omega_{\rm p} = 5^{\circ}$ and $\omega_{\rm bin}=\omega_{\rm p}=\theta_{\rm bin}=\theta_{\rm p}=0^{\circ}$. The binary and planet eccentricities are initially set to either 0 or 0.4, and the four simulations are the different combinations of these eccentricities, shown in different colours outlined in the top left plot. In the bottom right plot is a zoomed version of the evolution of $I_{\rm p}$ with vertical lines denoting two windows of transitability for each simulation. Note that for the curves on the bottom row the two red curves overlap and the two blue curves overlap as the orbital precession is a function of $e_{\rm bin}$ but not $e_{\rm p}$. For this reason dashed lines are used.}
\label{fig:ecc_test}
\end{center}  
\end{figure*}

An additional example test was run to see the effect of eccentricity on time-dependent transitability. A base circumbinary system was created with $M_{\rm A}=1M_{\odot}$, $M_{\rm B}=0.5M_{\odot}$, $P_{\rm bin}=5$ d, $P_{\rm p} = 85$ d, $a_{\rm bin} = 0.065$ AU, $a_{\rm p} = 0.43$ AU, $I_{\rm bin} = 95^{\circ}$, $I_{\rm p}=100^{\circ}$, $\Omega_{\rm bin}=0^{\circ}$ and $\Omega_{\rm p} = 5^{\circ}$. Both $\omega$ and $\theta$ for the binary and planet were initially set to $0^{\circ}$. Four simulations were run with $e_{\rm bin}$ and $e_{\rm p}$ set to either 0 or 0.4. In Fig.~\ref{fig:ecc_test} is a plot of the variation of the orbital elements over time in the four simulations: both binary and planet orbits are initially eccentric (light blue), both are initially circular (light red), only the planet is eccentric (dark red) and only the binary is eccentric (dark blue). All plots have been normalised to the precession period. Detailing the complex orbital mechanics of circumbinary planets is beyond the scope of this work, and has been cover in various papers (e.g. \citealt{leung13,georgakarakos15}), so here only a brief description of each variation is provided.

\begin{itemize}

\item {\bf Eccentricity:} If the binary is circular then the planet, whether it be initially circular or initially eccentric, has a constant eccentricity. Conversely, an eccentric binary induces periodic variations in the planet's eccentricity of $\sim 0.05$ in magnitude. 
\item {\bf Semi-Major Axis:} In all four cases the planet's semi-major axis does not vary by more than $\sim 5\%$. As both planet and binary eccentricities increase, the variation in $a_{\rm p}$ increases.
\item {\bf Argument of Periapse:} For initially eccentric planets $\omega_{\rm p}$ has a simple behaviour between $0^{\circ}$ and $360^{\circ}$, although if the binary is eccentric there is some slight non-linearity and the apsidal advance of $\omega_{\rm p}$ is slightly faster than the precession of $\Omega_{\rm p}$. If both the planet and binary are circular then $\omega_{\rm p}$ is essentially undefined, leading to the light red fuzz covering most of the plot. For an initially circular planet around an eccentric binary $\omega_{\rm p}$ is initially undefined but becomes defined as $e_{\rm p}$ grows above zero under influence from the binary.
\item {\bf Longitude of the Ascending Node:} The behaviour of $\Omega_{\rm p}$ is dependent on $e_{\rm bin}$ and not $e_{\rm p}$. Consequently, the light and dark red curves are overlapping (circular binaries) and the light and dark blue curves are overlapping (eccentric binaries). For an eccentric binary the sinusoid is slightly skewed to the right, indicative of the variable precession rate that \citet{farago10} and \citet{doolin11} discovered.
\item {\bf Inclination:} The most important parameter for transitability is $I_{\rm p}$. Similar to $\Omega_{\rm p}$, the variation of $I_{\rm p}$ only depends on $e_{\rm bin}$ and not $e_{\rm p}$. For eccentric binaries (light and dark blue curves) we see that a greater range of $I_{\rm p}$ is covered. This is consistent with the tests of \citet{martin15} which showed eccentricity generally boosts the amount of planets in transitability.
\end{itemize}

The bottom right plot in Fig.~\ref{fig:ecc_test} is the variation of $I_{\rm p}$ zoomed to the windows of transitability\footnote{So that eccentricity can be accounted for, these are calculated using N-body simulations and numerical tests of overlapping orbits.}, demarcated by vertical lines of the corresponding colour. There are two such windows, one centred around $0.5T_{\rm prec}$ and one to the right. Since transitability occurs when $I_{\rm p}$ is near $90^{\circ}$, we see that the windows are somewhat similar for the first window of transitability. For the second window of transitability they are spread out by $\sim 10\%$ of the precession period. The windows of transitability tend to be longer for eccentric binaries (light and dark red curves) as a result of the smaller variation of $I_{\rm p}$, as shown in Sect.~\ref{subsec:percentage}.

A preliminary conclusion is that the predominant  effect of eccentricity on transitability is not how it changes the geometry but rather how it changes the orbital precession. In that sense, $e_{\rm p}$ has negligible effect as the evolution of $I_{\rm p}$ is unaffected by it, whilst $e_{\rm bin}$ may be important. There is also evidence that eccentricity increases transit probabilities but this is yet to be fully quantified.

\section{Applications to the known {\it Kepler} circumbinary planets}\label{sec:applications}

\begin{table*}
\caption{Orbital parameters of the transiting systems discovered so far by {\it Kepler}.} 
\centering 
\begin{tabular}{ ccccccccccccccccccccc } 
\hline\hline 
Name & $M_{\rm A}$ & $M_{\rm B}$ & $R_{\rm A}$ & $R_{\rm B}$ & $a$ & $P$ & $e$ & $I$ & $\Omega$ & $\omega$ & $\lambda$ & $\Delta I$ & Epoch \\ 
[0.5ex] 
& ($M_{\odot}$) & ($M_{\odot}$) & ($R_{\odot}$) & ($R_{\odot}$) & (AU) &  (day) &   & (deg) & (deg) & (deg) & (deg) & (deg) & (BJD)\\
\hline
\hline
16 Binary& 0.690 & 0.203 & 0.649 & 0.226 & 0.224 & 41.079 & 0.159 & 90.340 & 0 & 263.464 & 92.352 & - &  \multirow{ 2}{*}{2,455,212.123}\\
16 Planet b& - & - & - & - & 0.705 & 228.776 & 0.007 & 90.032 & 0.003 & 318 & 106.51 & 0.308 &  \\
\hline
34 Binary& 1.048 & 1.021 & 1.162 & 1.093 & 0.229 & 27.796 & 0.521 & 89.858 & 0 & 71.436 & 300.197 & - & \multirow{ 2}{*}{2,454,969.200} \\
34 Planet b& - & - & - & - & 1.090 & 288.822 & 0.182 & 90.355 & -1.74 & 7.907 & 106.5 & 1.810 & \\
\hline
35 Binary& 0.888 & 0.809 & 1.028 & 0.786 & 0.176 & 20.734 & 0.142 & 90.424 & 0 & 86.513 & 89.178 & -& \multirow{ 2}{*}{2,454,965.850}\\
35 Planet b& - & - & - & - & 0.604 & 131.458 & 0.042 & 90.76 & -1.24 & 64.093 & 136.4 & 1.285 & \\
\hline
38 Binary& 0.949 & 0.249 & 1.757 & 0.272 & 0.147 & 18.795 & 0.103 & 89.265 & 0 & 268.680 & 236.733 &  -&  \multirow{ 2}{*}{2,454,970.0}\\
38 Planet b& - & - & - & - & 0.464 & 105.595 & 0.032 & 89.446 & -0.012 & 32.829 & 37.817 & 0.181 & \\
\hline
47 Binary& 1.043 & 0.362 & 0.964 & 0.351 & 0.084 & 7.448 & 0.023 & 89.34 & 0 & 212.3  &  235.85 & -&  \multirow{ 3}{*}{2,455,000.0}\\
47  Planet b& - & - & - & - & 0.296  & 49.514 & 0.094 & 89.59 & 0.1 & 178.172 & 350.589 & 0.269 & \\
47 Planet c& - & - & - & - & 0.989 & 303.158 & 0.423 & 89.826 & 1.06 & 214.104 & 305.164 & 1.166  & \\
\hline
64 Binary& 1.384 & 0.336 & 1.734 & 0.378 & 0.174  & 20.000 & 0.212 & 87.360 & 0 & 217.6  & 291.6 &  -&  \multirow{ 2}{*}{2,454,900.0}\\
64 Planet b& - & - & - & - & 0.634 & 138.506 & 0.054 & 90.022 & 0.89 & 348.0 & 186.90 & 2.807 & \\
\hline
413 Binary& 0.820 & 0.542 & 0.776 & 0.484 &  0.101 & 10.116 & 0.037 & 87.322 & 0 & 279.74  &  62.887 & -&  \multirow{ 2}{*}{2,455,000.0}\\
413 Planet b& - & - & - & - & 0.355 & 66.262 & 0.118 & 89.929 & 3.139 & 94.6 & 0.5 & 4.073 & \\
\hline
453 Binary& 0.934 & 0.194 & 0.833 & 0.214 & 0.185  & 27.322 & 0.052 & 90.266 & 0 & 263.049  & 72.241 & - &  \multirow{ 2}{*}{2,454,964.0}\\
453 Planet b& - & - & - & - & 0.788 & 240.503 & 0.036 & 89.443 & 2.103 & 185.149 & 299.039 & 2.298 & \\
\hline
1647 Binary& 1.221 & 0.968 & 1.790 & 0.966 & 0.278  & 11.259 & 0.160 & 87.916 & 0 & 300.544  & 31.716 & - &  \multirow{ 2}{*}{2,455,000.0}\\
1647 Planet b& - & - & - & - & 2.721  & 1107.592 & 0.058 & 90.097 & -2.039 & 155.046 & 94.3780 & 3.016& \\
\hline 
\multicolumn{14}{l}{\footnotesize{{\bf Refs:} \citet{doyle11,welsh12,welsh14,orosz12a,orosz12b,schwamb13,kostov13,kostov14,kostov16}.}}\\
\multicolumn{14}{l}{\footnotesize{{\bf Note:} Updated elements for Kepler-453 and -1647 provided by Veselin Kostov (priv. comm.).}}\\
\multicolumn{14}{l}{\footnotesize{{\bf Note:} $\lambda$ is the mean longitude.}}\\
\multicolumn{14}{l}{\footnotesize{{\bf Note:} Kepler-47d is excluded because it has not yet been published and lacks a value for $\Delta I$.}}\\
\multicolumn{14}{l}{\footnotesize{{\bf Note:} Kepler-64 is also known as PH-1, as it was discovered by the Planet Hunters consortium: \url{https://www.planethunters.org/}}}\\

\end{tabular}
\label{tab:KeplerDiscoveries}
\end{table*}

\subsection{Predicted future transits and observations with {\it TESS}, {\it CHEOPS} and {\it PLATO}}\label{subsec:kepler_planets}

To the interest of those wanting follow-up transit observations of the known {\it Kepler} transiting systems,  this theory is applied to some upcoming photometric space missions. Whilst transit follow-up is possible from the ground, it is hampered in two ways. First, the known circumbinary planets all have 50+ day periods, considered long for transit studies. This makes scheduling difficult, particularly for ground-based observations. Second, transit durations of circumbinary planets (equation derived in \citealt{kostov14}) may be be significantly longer than equivalent single-star transits, owing to the relative motion of the two stars, and hence may be longer than an observing night.

The upcoming photometric missions we apply our results to are listed below. The {\it K2} mission is not included as it has never re-observed the original {\it Kepler} field.

\begin{itemize}

\item {\bf TESS:} A desired launch date of August 2017, after which {\it TESS} will observe the southern hemisphere for one year before observing the northern hemisphere for one year, starting roughly August 2018. Within this year, the original {\it Kepler} field is likely to receive roughly  one or two months of time coverage.

\item {\bf CHEOPS:} A desired launch date of December 2017 and a nominal 3.5 yr mission. Unlike the other missions, {\it CHEOPS} is not a transit survey for new planets but primarily a follow-up photometric space mission to better characterise known planets. In its low-Earth orbit, it has optimal observability of stars near the ecliptic. The {\it Kepler} field is observable but near the limit.

\item {\bf PLATO:} A desired launch date of early 2024 and a nominal 6 yr mission. The schedule of {\it PLATO} is unlikely to be decided until just a few years before launch, and will likely consist of many short observing fields, running for a few months, and one or two extended views, running for a one or multiple years. It is almost certain that the {\it Kepler} field will be re-observed at some time but it is not known when and for how long.

\end{itemize}

The orbital parameters of the known circumbinary planets are listed in Table~\ref{tab:KeplerDiscoveries}. For each system we calculate $I_{\rm p}(t)$ and the limits of transitability, $\left. I_{\rm p}\right|_{\rm transitability}$. The timing of actual transits on the primary and secondary stars is calculated using an N-body code. In light of the slight errors in the analytically calculated precession period (Sect.~\ref{subsec:precession_accuracy}), we use here the ``true" precession period which is taken from the N-body simulation. In Figs.~\ref{fig:Kepler_OverTime_One} and ~\ref{fig:Kepler_OverTime_Two} we show our results over 20 years between 2013 and 2033. This timespan covers the end of the original {\it Kepler} mission up until a few years past the future {\it PLATO} mission.

\begin{figure*}  
\begin{center}  
	\begin{subfigure}[b]{0.48\textwidth}
		\caption{\Large Kepler-16}
		\includegraphics[width=\textwidth]{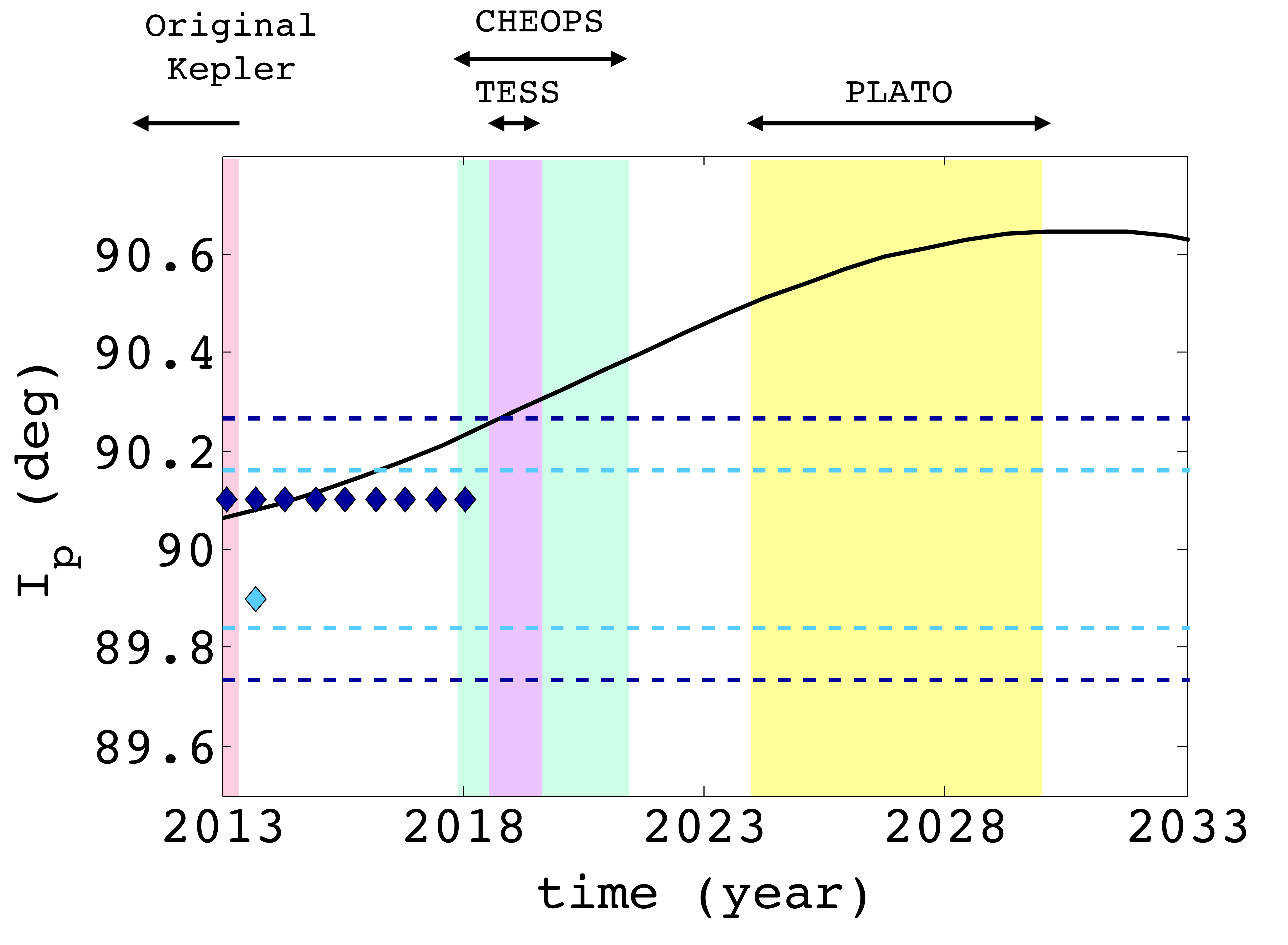}  
	\end{subfigure}
	\begin{subfigure}[b]{0.48\textwidth}
		\caption{\Large Kepler-34}
		\includegraphics[width=\textwidth]{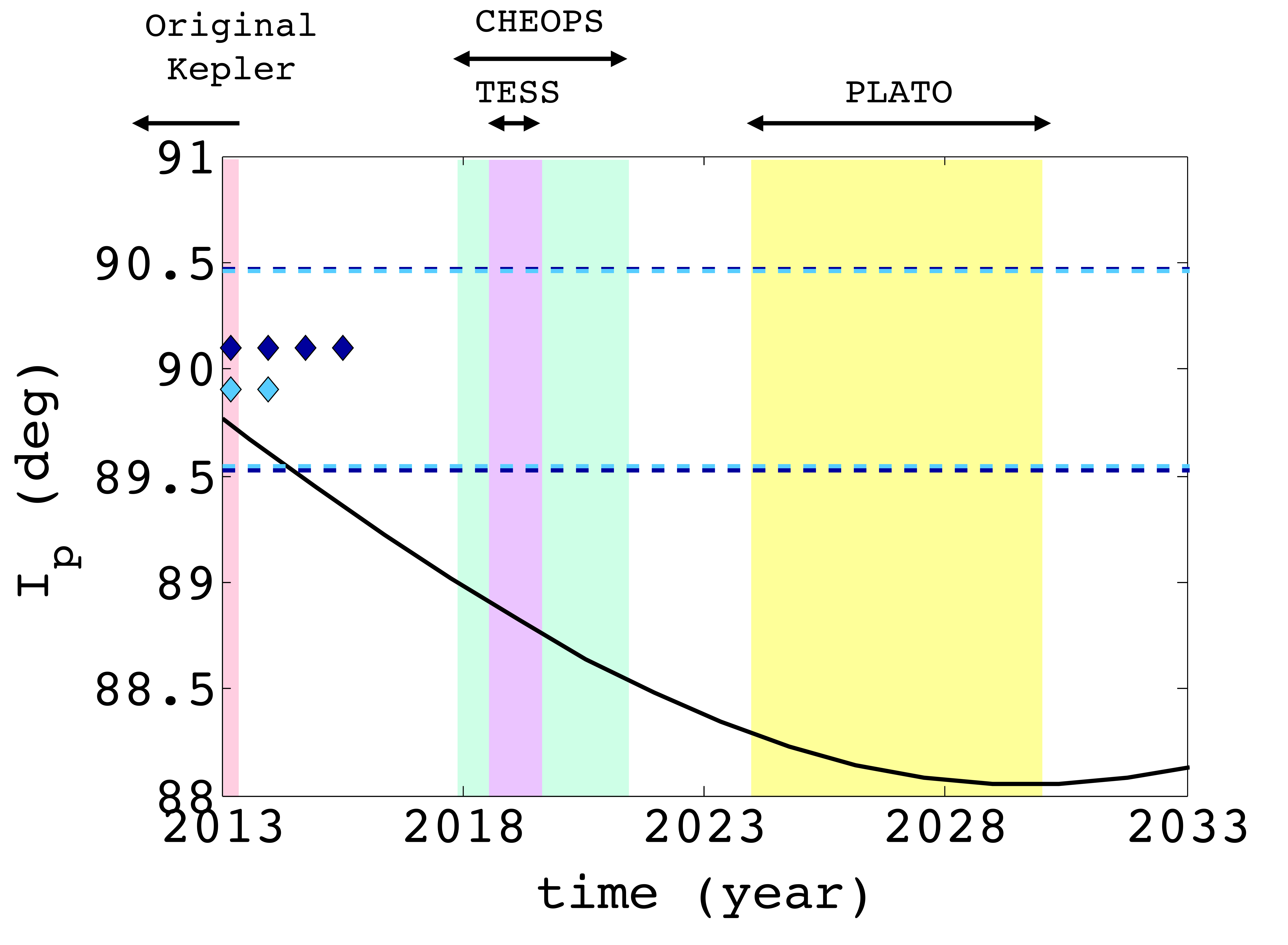}  
	\end{subfigure}
	\par\bigskip\bigskip
	\begin{subfigure}[b]{0.48\textwidth}
		\caption{\Large Kepler-35}
		\includegraphics[width=\textwidth]{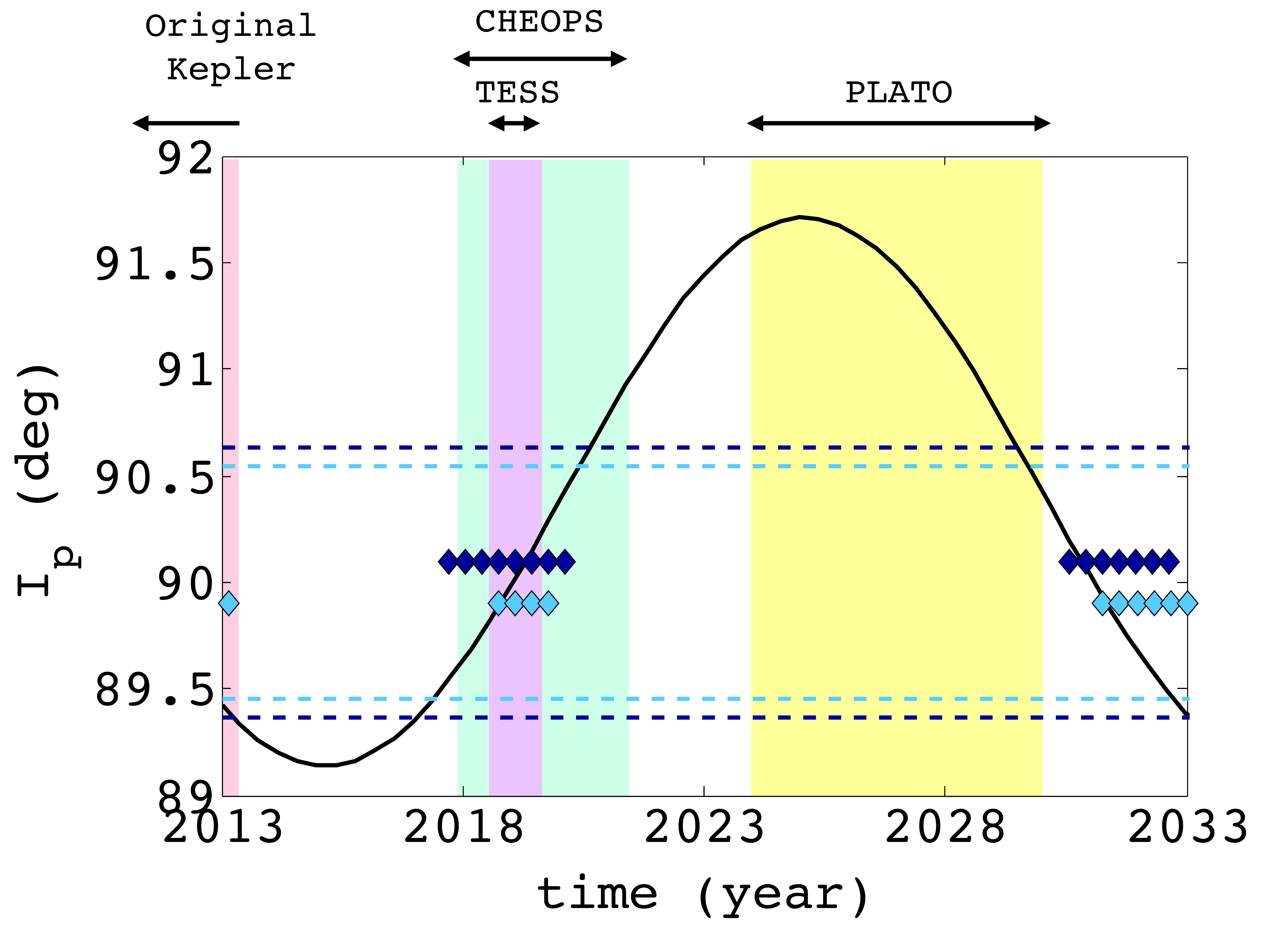}  	
	\end{subfigure}
		\begin{subfigure}[b]{0.48\textwidth}
		\caption{\Large Kepler-38}
		\includegraphics[width=\textwidth]{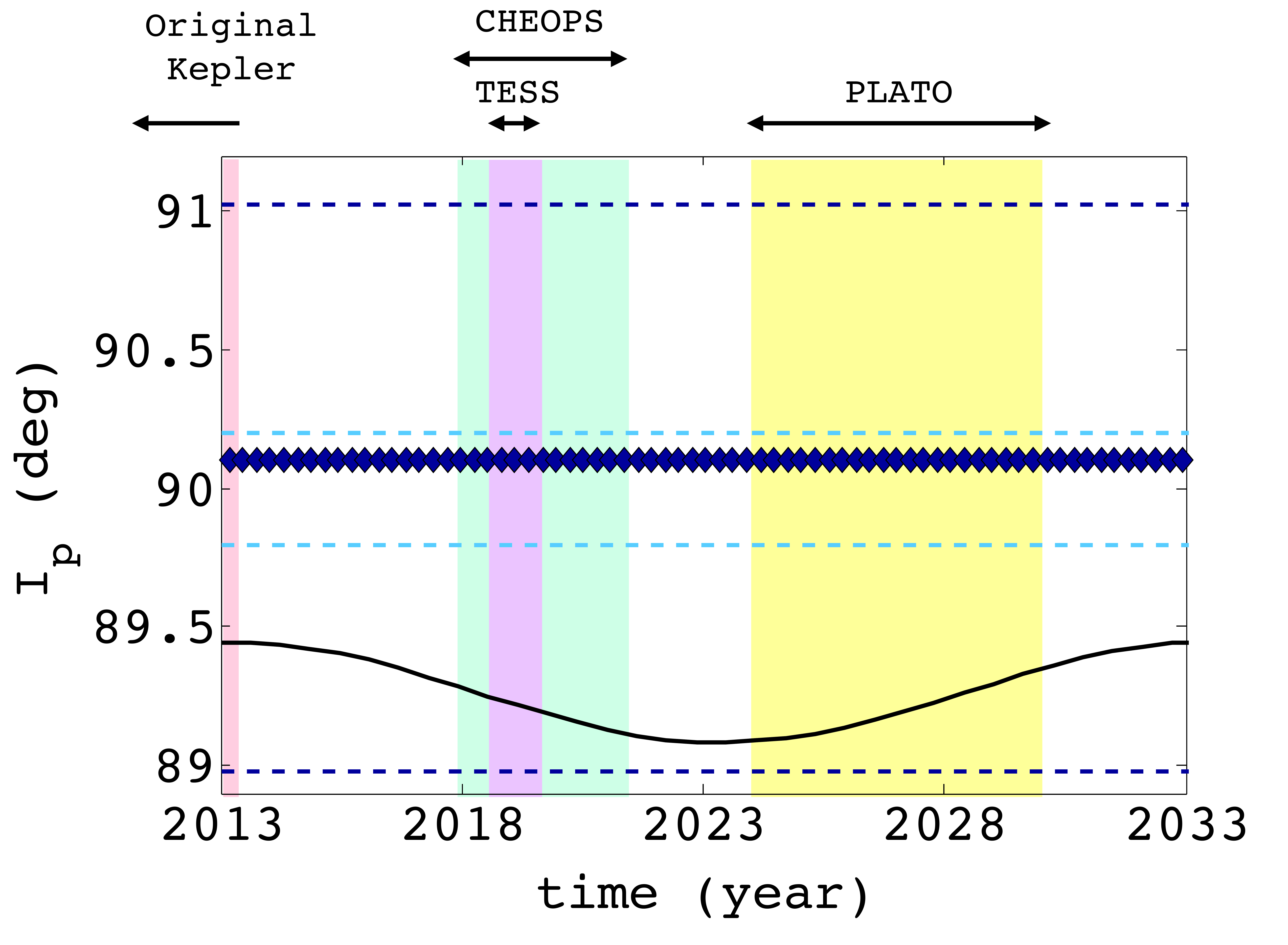}  
	\end{subfigure}
	\par\bigskip\bigskip
	\begin{subfigure}[b]{0.48\textwidth}
		\caption{\Large Kepler-47b}
		\includegraphics[width=\textwidth]{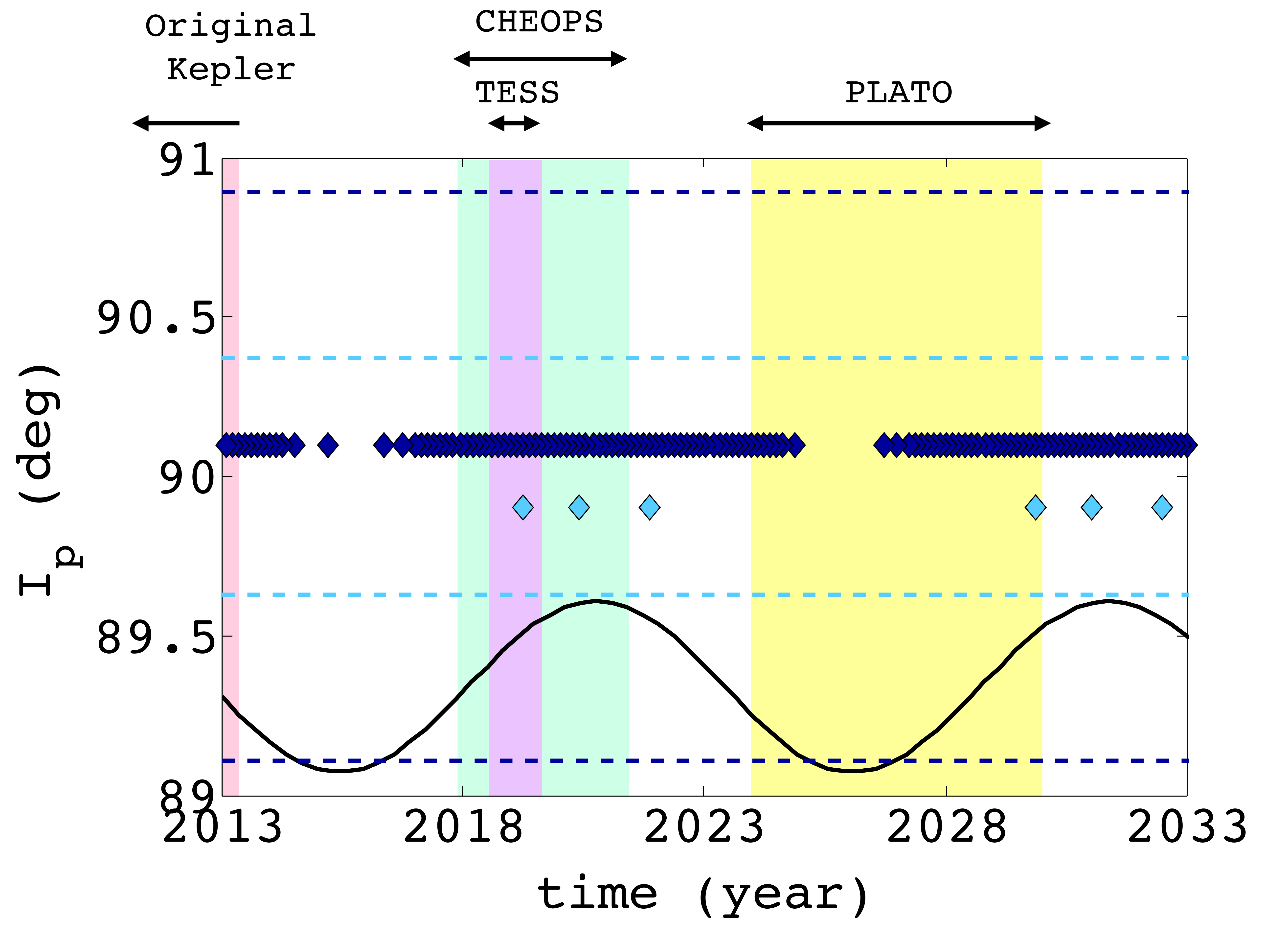}  
	\end{subfigure}
	\begin{subfigure}[b]{0.48\textwidth}
		\caption{\Large Kepler-47c}
		\includegraphics[width=\textwidth]{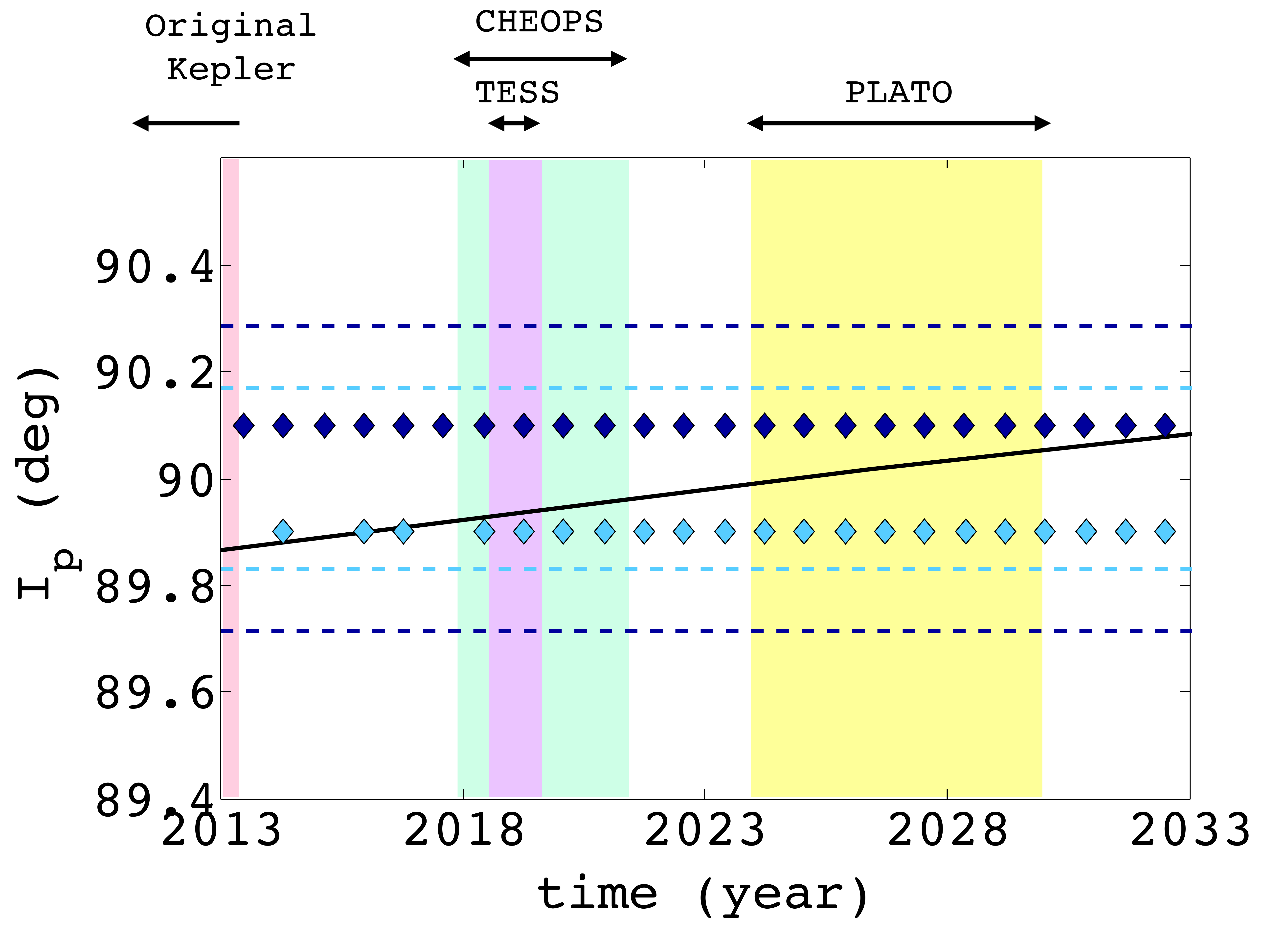}  	
	\end{subfigure}
		\caption{Time evolution of the planet inclination, $I_{\rm p}(t)$ (black curve) for known transiting circumbinary planets. The limiting inclinations for transitability (Eq.~\ref{eq:Ip_transitability}) are shown as horizontal dashed lines for the primary star (dark blue) and secondary star (light blue). Dark and light blue diamonds denote predicted transits on the primary and secondary star, respectively, calculated using an N-body simulation.  The vertical position of the diamonds has no physical meaning. Different coloured vertical bands denote the observing windows of different space telescopes, which we label above. Note that the {\it TESS} timespan commences after the {\it CHEOPS} timespan because even though {\it TESS} will launch beforehand, it will only observe the northern hemisphere in its second year. 
}\label{fig:Kepler_OverTime_One}  
\end{center}  
\end{figure*}

\begin{figure*} 
\ContinuedFloat
    \captionsetup{list=off,format=cont} 
\begin{center}  
	\begin{subfigure}[b]{0.48\textwidth}
		\caption{\Large Kepler-64}
		\includegraphics[width=\textwidth]{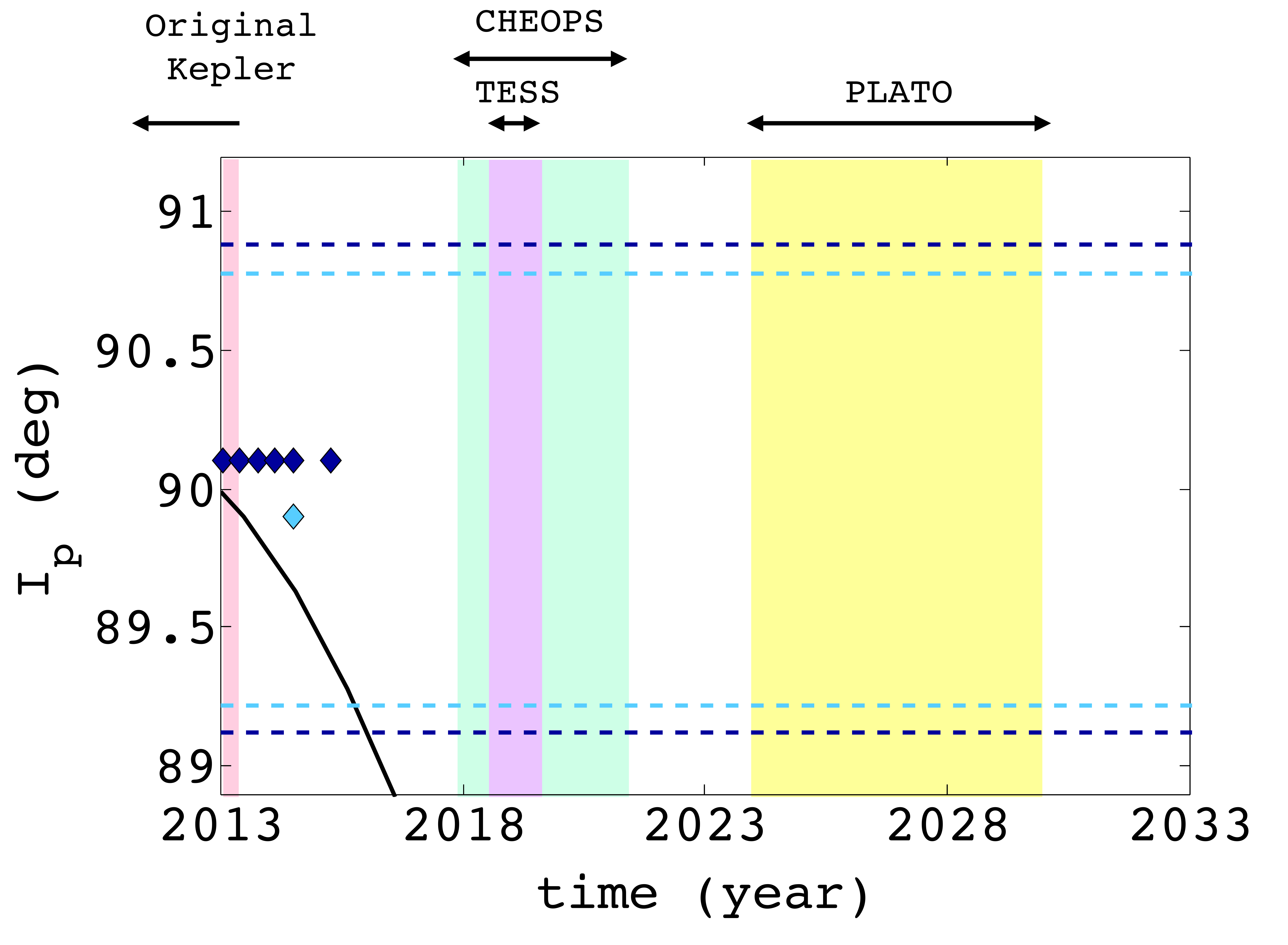}  
	\end{subfigure}
	\begin{subfigure}[b]{0.48\textwidth}
		\caption{\Large Kepler-413}
		\includegraphics[width=\textwidth]{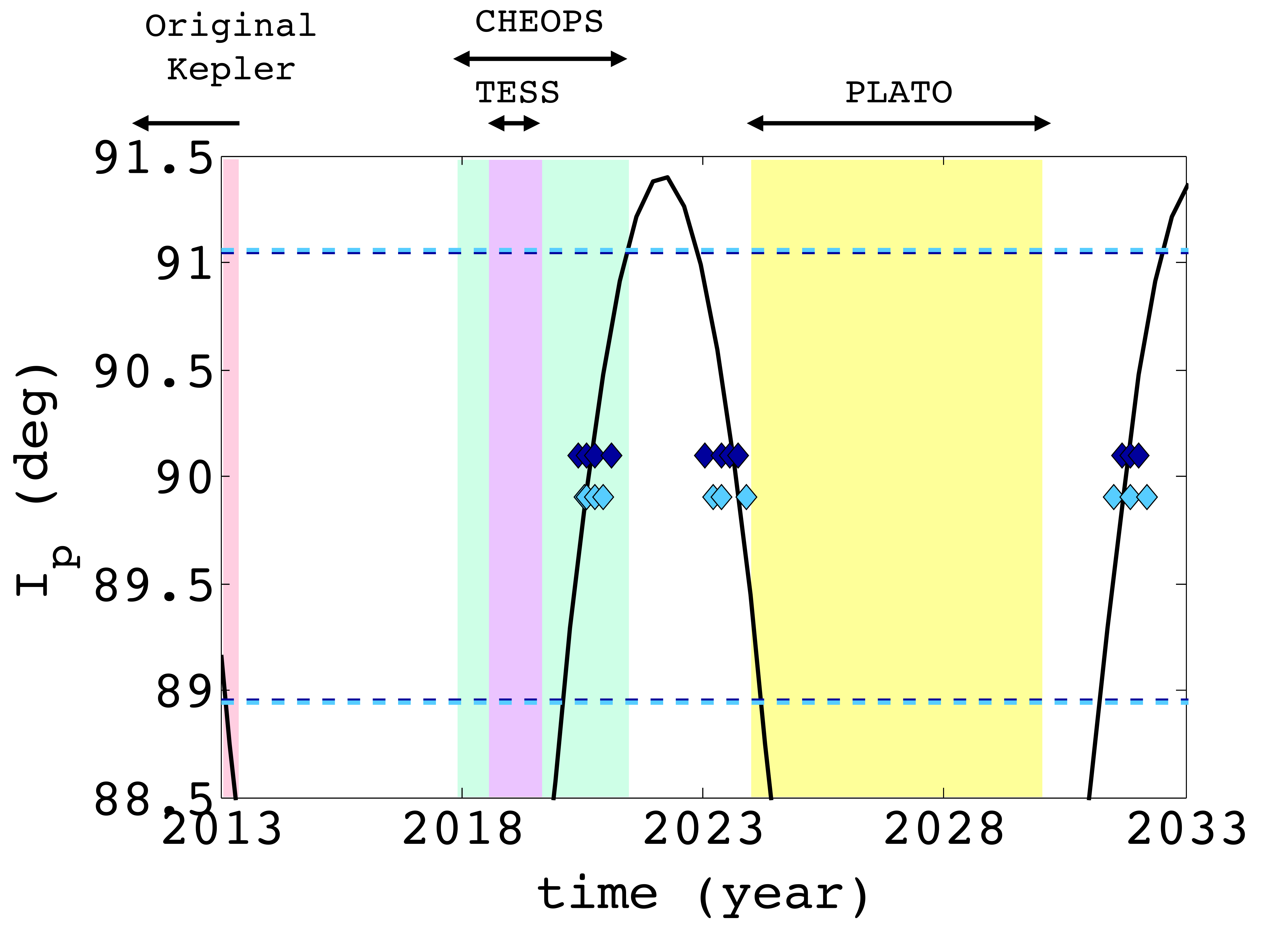}  
	\end{subfigure}
	\par\bigskip\bigskip
	\begin{subfigure}[b]{0.48\textwidth}
		\caption{\Large Kepler-453}
		\includegraphics[width=\textwidth]{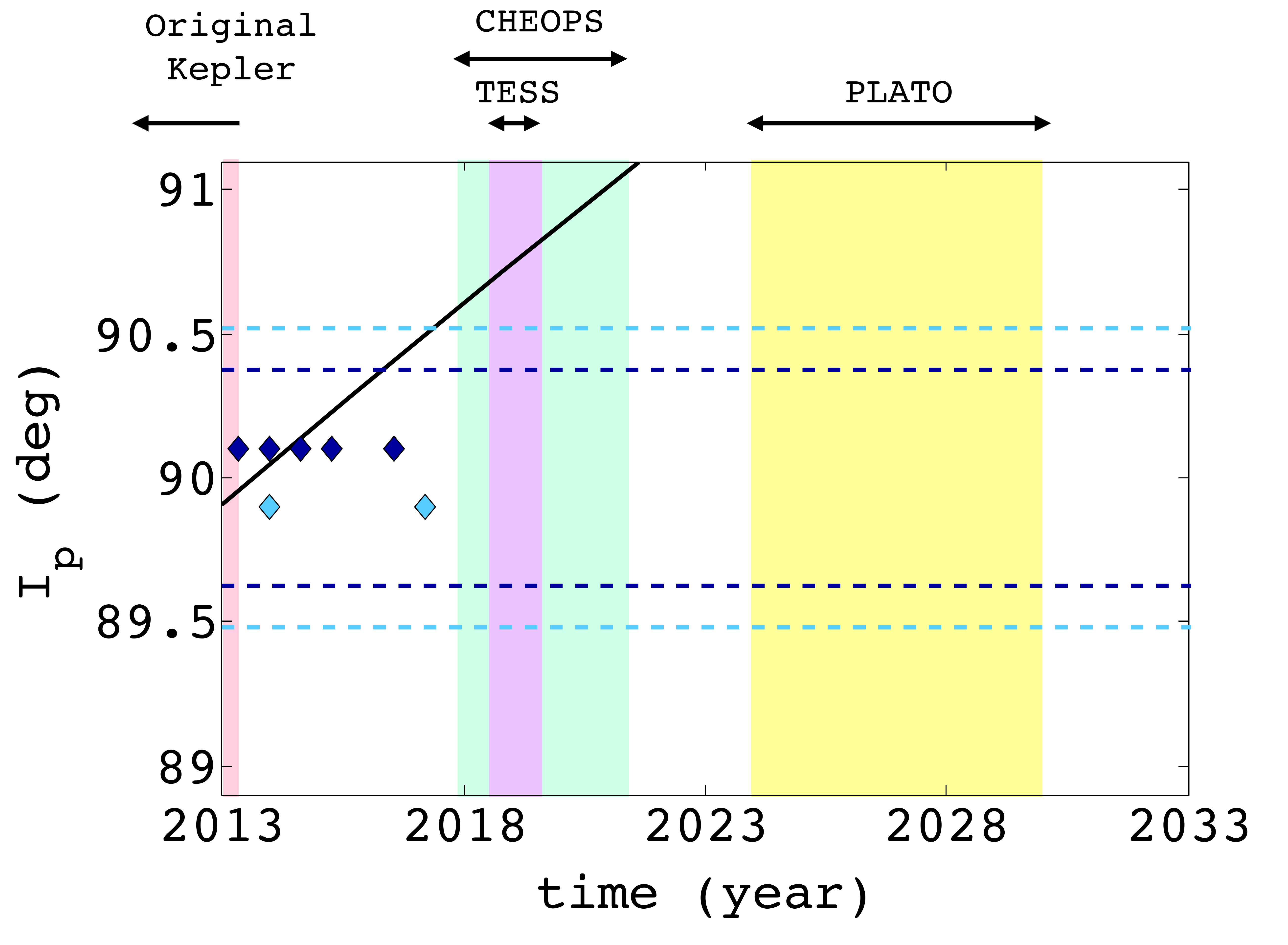}  	
	\end{subfigure}
	\begin{subfigure}[b]{0.48\textwidth}
		\caption{\Large Kepler-1647}
		\includegraphics[width=\textwidth]{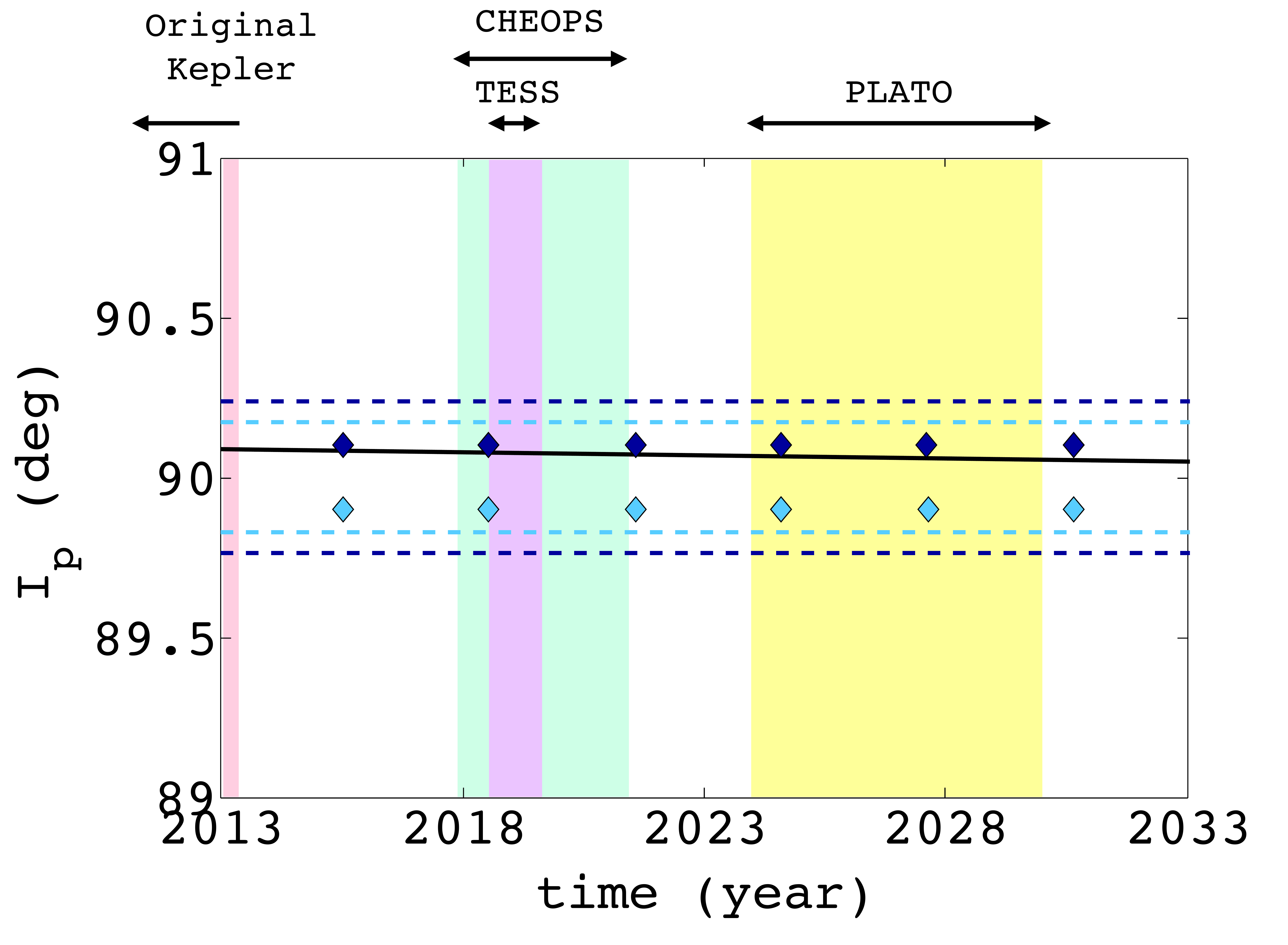}  	
	\end{subfigure}		
	\caption{Continued.}
\label{fig:Kepler_OverTime_Two}  
\end{center}  
\end{figure*} 

\begin{table*}
\caption{Observability of the known circumbinary planets discovered so far by {\it Kepler}  and predicted extra similar planets that will later transit. For observability a \cmark indicates analytic predicted transitability and in brackets are the number of transits predicted using N-body simulations}. 
\centering 
\begin{tabular}{l c c c c | c c c | c c c | c |} 
\hline\hline 
Name & \% primary  & \% secondary  & $T_{\rm prec}$ &  \multicolumn{3}{c}{Primary Observabiity}&  \multicolumn{3}{c}{Secondary Observability} & Predicted extra\\ 
 &  transitability &  transitability & (yr) &  {\it TESS} & {\it CHEOPS} & {\it PLATO} &  {\it TESS} & {\it CHEOPS} & {\it PLATO} & planets \\ 
[0.5ex] 
\hline 
16 & 42.5 & 30.3 & 41.4 & (0) & \cmark (1) & (0) & (0) & (0) & (0) & 0.9 \\
34& 16.9 & 16.4 &  69.3 &(0)  & (0) &(0) &(0) & (0) & (0) &3.4\\
35&36.0& 30.1 &  19.8& \cmark(3) & \cmark(7) & (0)&\cmark (3) & \cmark(4) & (0) &0.8\\
38& 100.0 & 0.0 & 20.3 & \cmark(3) & \cmark(13) & \cmark(21) & (0) & (0) & (0) & 0\\
47b& 82.5 & 0.0 & 10.6 & \cmark(7) & \cmark(27) & \cmark(30) & (1) & (2) & (1) & 0\\
47c& 19.7 &11.2 & 539.1 & \cmark(1)  & \cmark(4) & \cmark(7) & \cmark(1) & \cmark(4) & \cmark(7) & 3.9\\
64&28.4 & 26.9 & 35.4 &(0)  & (0) & (0)&  (0) & (0) & (0) & 1.5\\
413& 23.5 & 24.0 & 11.1 & (0) &\cmark (4) & (0)& (0) & \cmark(3) & (0) & 0.7\\
453& 10.5 & 14.6 & 102.9 & (0)& (0)& (0)& (0) & (0) & (0) & 6.0\\
1647& 7.0 & 5.1 & 7188.7 & (0)  & \cmark(1)& \cmark (2)& (0) & \cmark(1) & \cmark(2) &13.2\\
\hline 

\end{tabular}
\label{tab:Results}
\end{table*}

For each of the systems we summarise their observability in Table~\ref{tab:Results},  where a \cmark indicates transitability predicted by the analytic formula on the primary or secondary star and the number in brackets is the amount of transits found in the N-body simulation. The ``predicted extra planets" column is explained in Sect.~\ref{subsec:extra_planets}. Individual remarks on each system are provided below. 

\begin{itemize}

\item {\bf Kepler-16:} Unobservable except on the primary at the very start of {\it CHEOPS}, although this 2018 transit is predicted to be very grazing, with a duration of roughly one hour. 

\item {\bf Kepler-34:} The windows of transitability are nearly identical for the primary and secondary stars, owing to their similar mass and radius. The N-body code shows a transit in 2015 that occurs despite $I_{\rm p}(t)$ being outside the window of transitability. This is caused by the highly eccentric binary orbit ($e_{\rm bin}=0.52$),  which was shown in Sect.~\ref{subsec:eccentricity} to shift the windows of transitability. 

\item {\bf Kepler-35:} A good {\it CHEOPS} and {\it TESS} target, but illusive to the nominal {\it PLATO} mission.

\item {\bf Kepler-38:} The only known circumbinary planet with permanent transitability on the primary star, visible for all time. Contrastingly though, it will never transit the secondary star.

\item{\bf Kepler-47b:} The innermost planet in the three-planet system has primary transits that are fully observable by {\it TESS} and {\it CHEOPS} but a gap in transitability means that it could be missed by the {\it PLATO} mission. Analytically we predict the planet to be right near the edge of transitability on the secondary star. Numerically,  some transits across the secondary star  are predicted, however since the secondary star is significantly fainter than the primary such transits unlikely to be observable. The  primary transit signature of this  system is also a nice illustration that there is generally not a sharp boundary between transits occurring and not. When the planet exits transitability around 2014 and 2024, it does so ``gradually," meaning that it goes from transiting every passing to missing a few transits when the orbits are barely overlapping to having transits cease altogether. This is one of the complications that makes a precise analytic derivation of a time-dependent probability difficult.

\item {\bf Kepler-47c:} The outermost planet does not have permanent transitability but transits do not cease until well after the {\it PLATO} mission. For such a long period planet transitability is highly efficient, as explained in Sect.~\ref{subsec:percentage}. No predictions are made for the unpublished Kepler-47d, which is believed to reside between planets `b' and `c'.

\item {\bf Kepler-64/PH-1:} Transits ceased shortly after the end of the {\it Kepler} mission and will not return for decades.

\item {\bf Kepler-413:} The planet with the shortest precession period and largest misalignment generally only produces short bursts of 3-5 transits before an extended absence. It will be observable by {\it CHEOPS} but unfortunately not {\it TESS}, and will sneakily transit just before and after the {\it PLATO}.

\item {\bf Kepler-453:} With a similar misalignment to Kepler-413, this planet spends most of its time outside of transitability. However, with a much larger ratio of $T_{\rm p}/T_{\rm bin}$ and hence longer $T_{\rm prec}$, there is no chance of further observations for many decades.

\item {\bf Kepler-1647:}  One of the longest-period confirmed transiting planet, around one or two star(s), takes over 7,000 years to precess, and hence its orbit is essentially static within the next few decades. Its very long orbital period makes transitability highly efficiency, but the downside is that its period is similar to the mission lengths, and hence will be missed by {\it TESS} and only  six primary and secondary transits are visible by {\it CHEOPS} and {\it PLATO} combined. It also spends very little of its precession period within transitability.
\end{itemize}

\begin{figure}  
\begin{center}  
\includegraphics[width=0.49\textwidth]{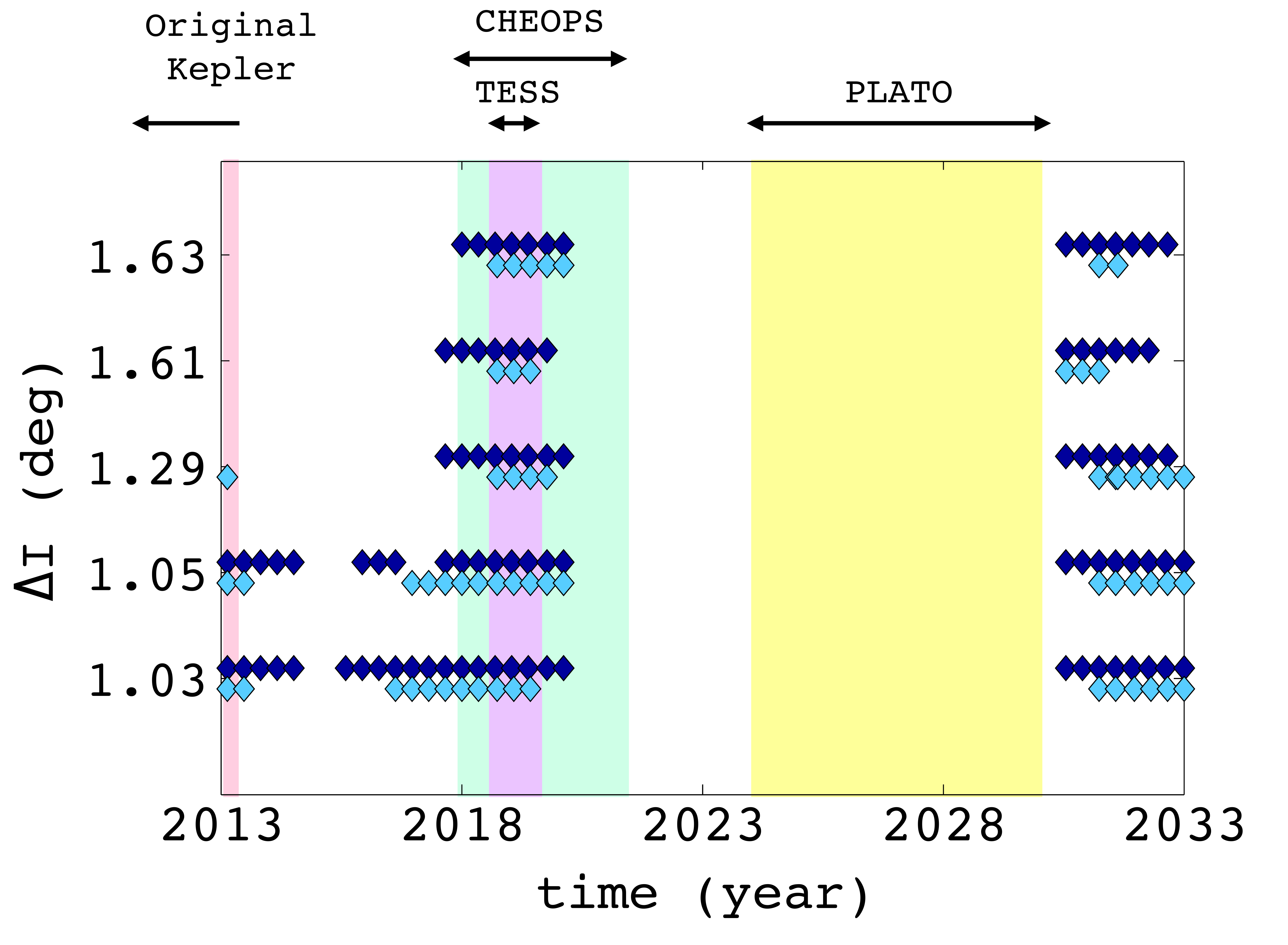}  
\caption{ Predicted primary (dark blue diamonds) and secondary (light blue diamonds) for Kepler-35 as a function of the mutual inclination. The five values of $\Delta I$ are drawn from the $2\sigma$ uncertainty bounds of $I_{\rm p}$ and $\Omega_{\rm p}$ published in \citet{welsh12}. From top to bottom, $\Omega_{\rm p} = $ -1.57$^{\circ}$, -1.57$^{\circ}$, -1.24$^{\circ}$, -1.0$^{\circ}$, -1.0$^{\circ}$ and $I_{\rm p} = $ 90.88$^{\circ}$, 90.76$^{\circ}$, 90.76$^{\circ}$, 90.76$^{\circ}$ and 90.67$^{\circ}$. Equation~\ref{eq:Delta_I} is used to calculate $\Delta I$. The middle row of data corresponds to the nominal values in Table.~\ref{tab:KeplerDiscoveries}.}
\label{fig:ObservationalUncertainty}
\end{center}  
\end{figure} 

Finally, is worth noting that the precision of predicted transits decreases the further one looks into the future. This is applicable to planets around  both one and two stars; errors in the ephemerides compound and the transit timing uncertainty may become longer than a typical transit duration.  For future characterisation say with the {\it James Webb Space Telescope} or the {\it European-Extremely Large Telescope}, the astronomical cost and competitiveness of these telescopes makes it impractical to have a very large transit window purely because of ``stale" ephemerides. 

The problem is amplified for circumbinary planets, owing to the high sensitivity of the transit timings as a function of the orbital parameters.  Uncertain ephemerides not only affect the timing of transits but whether or not they occur at all. To illustrate this effect, In Fig.~\ref{fig:ObservationalUncertainty} the primary and secondary transit times of Kepler-35 are shown for five different mutual inclinations which are all compatible with the $2\sigma$ errors published in \citet{welsh12}. The middle row of transit times are the same as in Fig.~\ref{fig:Kepler_OverTime_One}c. The number of predicted transits is a sensitive function of $\Delta I$,\footnote{Although it seems that {\it PLATO} has no chance of observing this target.} as was discovered in \citet{martin14} (Fig. 5 in that paper). Better predictions of future transit times requires a re-analysis using the full four years of {\it Kepler} data (Kepler-35 was published using 671 days of data).

\subsection{The number of similar planets await to be found orbiting the same {\it Kepler} eclipsing binaries}\label{subsec:extra_planets}

 Out of the ten published transiting circumbinary planets, how {\it lucky} were we to observe them? For a given set of binary and planet parameters, including inclinations, it is interesting to quantify how fortunate we were to have {\it Kepler's} four years of observations coincide with the window of transitability. From this, we can quantify the opposite case of being {\it unlucky} and missing transits, and hence we can estimate the amount of essentially identical circumbinary planets that may exist around eclipsing binaries discovered by {\it Kepler}, but have not {\it yet} transited.

Define $D$ as the probability of that a continuous observing campaign of length $T_{\rm obs}$ detects a planet transiting that spends $T_{\rm transitability}$ of its precession period in transitability,

\begin{equation}
\label{eq:discovery_probability}
D  = \min \left(\frac{T_{\rm obs} + T_{\rm transitability}}{T_{\rm prec}},1 \right).
\end{equation}
For simplicity, simply consider transitability on the primary star. Equation ~\ref{eq:discovery_probability} only works if the planet always transits a couple of times within transitability, in order to be detectable. For the low mutual inclination {\it Kepler} planets this is a valid assumption, as demonstrated in Fig.~\ref{fig:Kepler_OverTime_One}. The opposite probability of a failed detection, $F$, is simply 

\begin{equation}
\label{eq:fail_probability}
F  = \max \left(1-D,1 \right).
\end{equation}
We can therefore say that for a given system there should be $E$ extra circumbinary systems with essentially identical properties that will transit sometime in the future, and this is calculated as

\begin{align}
\label{eq:extra_planets}
E &= \frac{F}{D} \\
&= \max\left(\frac{T_{\rm prec}}{T_{\rm obs}+T_{\rm transitability}}-1,0\right).
\end{align}

Included in Table.~\ref{tab:Results} is the predicted number of extra planets, where $T_{\rm obs} = 4$ yr. I highlight here a couple of examples. For Kepler-16 $E=0.9$, which means we are essentially missing another Kepler-16-like planet that will transit in the future. For Kepler-38 $E=0$ because the planet has permanent transitability on the primary star and hence cannot evade detection. For Kepler-47b $E=0$ also, because not only does it spend a large percentage of its time in transitability but its precision period is only 10.6 yr, the shortest of all known circumbinary planets. At the other extreme, Kepler-453 and -1647 have $E=6.0$ and $13.2$, respectively, owing to long precision periods (particularly for 1647) and short percentages of transitability.

In total these simple estimates predict $\sim 30$ essentially identical circumbinary planets to ultimately transit {\it Kepler} eclipsing binaries. The number is reduced to 17 if Kepler-1647 is excluded, for which the wait time may be thousands of years.

\section{Conclusion}
\label{sec:conclusion}

We have derived analytic criteria for the time-dependence of transitability, a state where the planet and binary orbits intersect on the plane of the sky, which is a necessary but not sufficient condition for circumbinary transits. Equations calculated in this paper are applicable to both eclipsing and non-eclipsing binaries and planets of any mutual inclination. This paper improves upon the time-independent criteria derived in \citet{martin15}, and is a key step towards a complete analytic time-dependent transit probability. By calculating future transits of the 10 published transiting circumbinary planets, we predict that 4 may be observable by {\it TESS}, 7 by {\it CHEOPS} and 4 by {\it PLATO}. Interestingly, most of the planets spend less than 50\% of their time in transitability, some as low as $\sim$ 10-20 \%. As a consequence, there are likely $\sim 17-30$ circumbinary planets around binaries in the eclipsing binary catalog, that have not yet precessed into view. Such new planets may be revealed by the future {\it TESS} and {\it PLATO} surveys, or complementary methods such as eclipse timing variations.

\section{Acknowledgements}
A special thank you to Amaury Triaud, with whom I started this project and will ultimately finish it! I have also benefited from constant support by my PhD supervisor Stephane Udry.  Thank you to Veselin Kostov who is a dead-set ledge for helping out with some of the orbital elements of the known systems. This work was greatly aided by fruitful conversations with Rosemary Mardling and Javiera Rey. Finally, I thank the anonymous referee for useful suggestions that helped improve the paper, in particular motivating a deeper look into the effects of eccentricity. I also made an extensive use of \href{http://adsabs.harvard.edu}{ADS}, \href{http://arxiv.org/archive/astro-ph}{arXiv} and the two planets encyclopaediae \href{exoplanet.eu}{exoplanet.eu} and \href{exoplanets.org}{exoplanets.org} and thank the teams behind these online tools, which greatly simplify the research.

\end{document}